\documentclass[12pt]{article}
\pdfoutput=1
\usepackage{microtype}
\usepackage[english]{babel}
\usepackage[utf8]{inputenc}

\usepackage{hyperref}
\usepackage{lipsum}
\usepackage{curve2e}
\definecolor{gray}{gray}{0.4}

\usepackage{graphicx}
\usepackage{bm}
\def\be{\begin{equation}}
\def\ee{\end{equation}}
\def\bea{\begin{eqnarray}}
\def\eea{\end{eqnarray}}
\usepackage{latexsym}
\usepackage{dcolumn}
\usepackage{epsfig,amssymb,euscript}
\usepackage{amsmath}
\usepackage{array,calc,epsfig}
\usepackage{braket}
\usepackage{appendix}
\usepackage{bbold}
\usepackage{slashed}
\usepackage{verbatim}
\usepackage[table]{xcolor}
\usepackage{multirow}

\usepackage{subcaption}

\usepackage{chngcntr}
\usepackage[noadjust]{cite}

\usepackage{color}

\newcommand{\Tr}{{\rm Tr}}

\newcommand{\w}{\wedge}
\renewcommand{\to}{\rightarrow}
\renewcommand{\l}{\lambda}

\def\be{\begin{equation}}
\def\ee{\end{equation}}
\def\ba{\begin{eqnarray}}
\def\ea{\end{eqnarray}}
\def\bm{\begin{bmatrix}}
\def\em{\end{bmatrix}}
\def\bpm{\begin{pmatrix}}
\def\epm{\end{pmatrix}}
\def\nb{\nonumber}
\def\p{\partial}

\def\a{\alpha}

\def\e{\epsilon}

\def\g{\gamma}
\def\G{\Gamma}
\def\d{\delta}
\def\D{\Delta}
\def\l{\lambda}

\def\m{\mu}
\def\n{\nu}

\def\r{\rho}
\def\s{\sigma}

\def\t{\tau}

\def\le{\left}

\def\pri{\prime}
\def\k{\kappa}
\def\le{\langle}
\def\re{\rangle}
\def\bml{\begin{multline}}
\def\eml{\end{multline}}

\def\mc{\mathcal}

\def\ms{\mathsf}

\counterwithin*{equation}{section}

\begin{document}
\baselineskip=15.5pt
\pagestyle{plain}
\setcounter{page}{1}
\newfont{\namefont}{cmr10}
\newfont{\addfont}{cmti7 scaled 1440}
\newfont{\boldmathfont}{cmbx10}
\newfont{\headfontb}{cmbx10 scaled 1728}
\renewcommand{\theequation}{{\rm\thesection.\arabic{equation}}}
\renewcommand{\thefootnote}{\arabic{footnote}}
\vspace{1cm}
\begin{titlepage}
\vskip 2cm
\begin{center}
{\Large{\bf Resonance contributions to nucleon spin structure\\ in Holographic QCD}}
\end{center}
\vskip 10pt
\begin{center}
Francesco Bigazzi$^{a}$, Federico Castellani$^{a,b}$
\end{center}
\vskip 10pt
\begin{center}
\vspace{0.2cm}
\textit {$^a$ INFN, Sezione di Firenze; Via G. Sansone 1; I-50019 Sesto Fiorentino (Firenze), Italy.}\\
\textit{$^b$ Dipartimento di Fisica e Astronomia, Universit\'a di Firenze; Via G. Sansone 1;\\ I-50019 Sesto Fiorentino (Firenze), Italy.}
\vskip 20pt
{\small{bigazzi@fi.infn.it, federico.castellani@unifi.it}
}

\end{center}

\vspace{25pt}

\begin{center}
 \textbf{Abstract}
\end{center}

\noindent 
We study polarized inelastic electron-nucleon scattering at low momentum transfer in the Witten-Sakai-Sugimoto model of  holographic QCD, focusing on resonance production contributions to the nucleon spin structure functions. Our analysis includes both spin $3/2$ and spin $1/2$ low-lying nucleon resonances with positive and negative parity. We determine, in turn, the helicity amplitudes for nucleon-resonance transitions and the resonance contributions to the neutron and proton generalized spin polarizabilities. Extrapolating the model parameters to realistic QCD data, our analysis, triggered by recent experimental results from Jefferson Lab, agrees with the observation that the $\Delta(1232)$ resonance gives the dominant contribution to the forward spin polarizabilities at low momentum transfer. The contribution is negative and tends to zero as the momentum transfer increases. As expected, the contribution of the $\Delta(1232)$ to the longitudinal-transverse polarizabilities is instead negligible. The latter, for both nucleons, turn out the be negative functions with zero asymptote. The holographic results, at least for the proton where enough data are available, are in qualitative agreement with the resonance contributions to the spin polarizabilities extracted from experimental data on the helicity amplitudes. 

\end{titlepage}

\tableofcontents

\section{Introduction}\label{intro}
The analysis of nucleons' internal structure is an exciting subject which still provides puzzles and challenges to both theoretical and experimental physics. Among the various observables under  scrutiny, a special place is occupied by the so-called polarizabilities. They describe the collective response of a composite object to external electromagnetic fields. The distortion of charge and magnetization distributions is described by the electric and magnetic polarizabilities. Spin polarizabilities, instead, quantify the spin-dependent internal rearrangement of polarization densities. In the case in which the probing processes involve virtual photons, with squared momentum transfer $Q^2>0$, the above observables are called generalized polarizabilities. 

Spin observables, which are the main focus of the present work, yield a wealth set of tests to our current understanding of the nucleon internal structure. Generalized spin polarizabilities are determined studying inelastic scattering processes involving polarized electrons and polarized nucleon targets: changing the relative orientation between nucleon and electron spins one extracts the related cross sections and then the polarizabilities. These are the forward (transverse-transverse) spin polarizability $\gamma_0(Q^2)$ and the longitudinal-transverse spin polarizability $\delta_{LT}(Q^2)$. Technically, they are related to moments of certain combinations of the nucleon polarized spin structure functions. See for instance the excellent reviews \cite{Jaffe:1996zw, Deur:2018roz, Manohar:1992tz,Drechsel:2002ar}.

In a couple of recent publications \cite{JeffersonLabE97-110:2019fsc,E97-110:2021mxm} (see also \cite{Deur:2022sot} for a review), the Jefferson Lab E97-110 Collaboration has presented novel results on the behavior of $\gamma_0(Q^2)$ and $\delta_{LT}(Q^2)$ for the neutron, at very low energy-momentum transfer squared (namely $Q^2$ down to 0.035 $\rm{GeV}^2$ with beam energies ranging between $1.1$ and $4.4\,{\rm GeV}$). At low energies, in the resonance region far below the deep-inelastic scattering domain, perturbative QCD cannot be applied and it is widely expected that sensible theoretical tools for extracting predictions on the above observables could be provided by chiral effective field theories (see for instance \cite{Bernard:2002pw, Kao:2002cp, Bernard:2012hb, Lensky:2014dda,Alarcon:2020icz}) or other phenomenological models (see e.g. \cite{Drechsel:2000ct}).\footnote{Lattice QCD calculations are still not very much under control for low $Q^2$ spin observables, see e.g. \cite{Deur:2018roz}.} However, experimental data from JLab show some substantial qualitative and quantitative disagreement with the latter, especially for what concerns the low $Q^2$ behavior of the longitudinal-transverse neutron spin polarizability. In particular, at small $Q^2$, neutron's $\delta_{LT}(Q^2)$  turns out to be small, even negative, with a positive slope, whilst theoretical predictions give a positive and decreasing function. This is the updated version of what is often referred to as the $\delta_{LT}$-puzzle.

The latter data/prediction discrepancy demanded to investigate if similar issues exist for the proton. The answer to this question has been given in 2022 by the JLab Hall A $g_2p$ Collaboration \cite{JeffersonLabHallAg2p:2022qap}, which has presented the first ever data for $\delta_{LT}(Q^2)$ for the proton for a range of $Q^2$ from $0.02$ to $0.13$ GeV$^2$ (with beam energies ranging between $1.2$ and $3.4\,{\rm GeV}$). The results show general agreement with theoretical predictions in \cite{Alarcon:2020icz}, while there is a discrepancy with those in \cite{Bernard:2002pw}. For what concerns the generalized forward spin polarizability $\gamma_0(Q^2)$ of the proton, novel results are shown in a recent article from CLAS \cite{CLAS:2021apd}. Once again this observable in the low energy regime does not display a behavior fully compatible with standard theoretical predictions.

It is worth mentioning that while processes involving nucleons and pion loops are quite under control in chiral effective field theories, accounting for the contributions of the production of heavier hadrons and in particular of baryonic resonances - starting from the most relevant one, i.e. the spin $3/2$ $\Delta(1232)$ - is typically difficult in general. Some interesting advance in the field can be found in e.g. \cite{Alarcon:2020icz} where the contributions to nucleon spin polarizabilities of both $\Delta$-exchange and of $\Delta$-pion loops are taken into account in a chiral effective model. Since resonance contributions are expected to be important at low energies, it is certainly useful to study these issues under the lens of complementary theoretical approaches.

Motivated by the above mentioned challenging experimental results, in this work we aim to begin a study of spin-dependent observables at low momentum transfer in a QCD-like model whose strong coupling regime can be analyzed by means of the holographic gauge/gravity correspondence. This non-perturbative tool allows to shed light into physical problems which are  sometimes hard to explore within more standard effective approaches. 

The model we focus on is the top-down Witten-Sakai-Sugimoto (WSS) model \cite{Witten:1998zw, Sakai:2004cn}. This is a QCD-like $SU(N_c)$ theory, with $N_f$ quarks\footnote{In this work we will focus on the $N_f=2$ case with zero quark masses.
} and massive adjoint matter fields. Crucially, the model displays confinement and chiral symmetry breaking at low energies. These features can be readily deduced, in the planar limit at strong coupling, by the dual holographic description. When $N_f\ll N_c$,  the latter displays $N_f$ extended sources (D8-branes) probing a specific curved classical gravity background. In the holographic limit, the low energy hadronic degrees of freedom of the field theory are described in terms of an effective five-dimensional $U( N_f)$ Yang-Mills-Chern-Simons theory in curved spacetime. 
In the WSS model, mesons are identified with modes of the five-dimensional gauge field, while baryons arise, as in the Skyrme model \cite{Adkins:1983ya}, as solitonic configurations (in this case instantons) \cite{Hata:2007mb}. Their static properties and elastic form factors have been studied in \cite{Hashimoto:2008zw}. Extrapolating the model parameters to QCD, many hadronic properties have been computed in the literature often with ${\cal O}(10\%-20\%)$ accuracy. 

In the present work we will provide a first explicit analysis of nucleon spin structure at low momentum transfer in the WSS model. In particular, we will study the contribution of the production of low-lying spin 3/2 and spin 1/2 baryonic resonances (with positive and negative parity), to the nucleon spin structure functions and to the above mentioned spin polarizabilities. The analysis involves the computation of  the so-called helicity amplitudes for the nucleon-to-resonance transitions. These are given in terms of  matrix elements of the electromagnetic current between resonance and nucleon states. 

Extrapolating the model parameters to realistic QCD data, our results indicate that the $\Delta(1232)$ resonance gives relatively large, negative contributions (which increase towards zero as $Q^2$ increases) to the generalized forward nucleon spin polarizabilities $\gamma_0(Q^2)$ at low $Q^2$. Its contribution to the longitudinal-transverse polarizabilities is instead negligible (actually zero, to leading order in the large $N_c$ holographic limit), as expected (see e.g. \cite{E97-110:2021mxm}). Different spin $1/2$ resonances give different contributions (in sign and amplitude) to  nucleons' $\delta_{LT}(Q^2)$. To leading order in the holographic limit, spin $1/2$ positive parity resonances do not contribute to neutron's $\delta_{LT}(Q^2)$ and give positive (and decreasing) contributions to proton's one. Negative parity ones, instead, give rise to negative contributions, going towards zero as $Q^2$ increases. For both nucleons, the net effect is to provide a negative $\delta_{LT}(Q^2)$ function, whose absolute value tends to zero as $Q^2$ increases. 

As we will see, the above features are in qualitative agreement with the resonance contribution to the spin polarizabilities as can be deduced (at least in the proton case, where enough data are available) from experimental results on the helicity amplitudes at low $Q^2$. This is certainly encouraging for what concerns the reliability of the holographic model. Still, this leaves the aforementioned puzzles unsolved, suggesting that non-resonant contributions, which are not taken into account in our analysis, play an important role. 

Previous interesting works \cite{Bayona:2011xj,Ballon-Bayona:2012txi} focusing on a class of spin $1/2$ resonances, accounted for resonance-proton transition form factors and resonance contributions to the unpolarized proton structure functions in the WSS model. 
Our analysis extends to all possible WSS spin $1/2$ resonances, includes the spin $3/2$ ones\footnote{Nucleon-to-Delta transition form factors in the WSS model, to leading order in the holographic limit, were considered in \cite{Grigoryan:2009pp}. In that work the spin-isospin structure of the baryonic current was neglected and a mesonic electromagnetic current was used instead.}, includes the neutron as possible target state, and it is meant to explore the polarized structure functions of both nucleons. 

This work is organized as follows. 

In section \ref{sec1} we provide a brief review of the basic observables we will be focusing on, from the hadronic tensor for lepton-nucleon scattering, to the related unpolarized and polarized structure functions. We introduce the helicity amplitudes for nucleon-resonance transitions and review how they provide the building blocks to account for resonance production contributions to the nucleon structure functions. Moreover, we introduce the generalized spin polarizabilities, given as moments of certain combinations of the polarized structure functions. Finally, we provide general expressions for sharp resonance contributions to the longitudinal-transverse (\ref{dlt3}) and the forward (\ref{god2}) generalized polarizabilities.

In section \ref{holorev} we summarize the main ingredients of the holographic description of hadrons in the WSS model, with a focus on baryonic states and electromagnetic current.

In section \ref{sec:helicity} we present our results for the nucleon-to-resonance helicity amplitudes in the WSS model, considering both spin $1/2$ and spin $3/2$ resonances with positive and negative parity.  

In section \ref{sec:compare}, we examine our results in more detail, providing, for some of the low-lying resonances on which we will focus on (namely the spin $3/2$ $\Delta(1232)$, the spin $1/2$ Roper $N(1440)$ and the negative parity spin $1/2$ $N(1535)$) explicit results for the helicity amplitudes and the related contribution to the nucleon spin polarizabilities. We also comment on the contributions from the spin $3/2$ $\Delta(1600)$, the spin $1/2$ $N(1710)$ and the negative parity $N(1650)$ resonances. 
Summing up the various contributions we will be able to provide an estimate of the total contribution of the above low-lying resonances to the nucleon spin polarizabilities at low-$Q^2$. Finally, we will compare our results with those obtained inferring the resonance contributions to proton's spin polarizabilities from experimental data on the helicity amplitudes.

We provide some concluding remark in section \ref{sec:conc}. Useful material can be found in the appendix.
%
%
\section{Nucleon structure functions}
\label{sec1}
Let us consider inclusive lepton-nucleon scattering
\be
\label{scattproc}
l\, N \rightarrow l'\, X\,,
\ee
where the hadronic final states $X$ are not observed and hence have to be summed over when computing the scattering cross section. 
Focusing on the case of electron scattering, the dominant interaction with the nucleon target consists in the one-photon exchange: if the target absorbs the (virtual) photon to produce a hadronic final state $X$, different from the initial nucleonic one, the scattering processes is said to be inelastic. 

In the laboratory frame, the initial electron with momentum $k^\m = (E,\vec k)$ exchanges a photon of momentum $q^\m$ with the target having a momentum $p^\m = (m,0)$. Being the process inclusive, only the outgoing electron with momentum $k^\prime= (E^\pri, \vec k^\pri)$ is detected. The kinematical variables relevant for this process are the photon virtuality\footnote{In this work we work with a mostly plus signature metric $(-,+,+,+)$. In the mostly minus convention, instead, $q^2=-Q^2<0$.}
\be
\label{virtuality}
q^2 \equiv (k-k')^2=  -(q^0)^2+ \vec q\,^2 = Q^2 >0\,,
\ee
and the Bjorken variable
\be
\label{x}
x= -\frac{Q^2}{2p\cdot q}= -\frac{Q^2}{2\n}\,, \hspace{1cm} \n = p\cdot q\,.
\ee
The invariant mass of the final hadronic state $X$ (with a total baryon number equal to that of the target) is given by
\be
m_X^2 = -(p+q)^2 = m^2 - 2p\cdot q -q^2,
\ee
with the constraint 
$ m_X^2 \geq m^2$, which implies $x\leq 1$. Moreover, since $Q^2>0$ and $\n$ is negative, the Bjorken variable $x$ must be positive, by definition (\ref{x}). Then, the physical region of scattering is $0<x\leq1$. In addition, the expression for $x$ can be rephrased as
\be
x = -\frac{Q^2}{2p\cdot q} = 1+ \frac{m_X^2 -m^2}{2p\cdot q} \to x= \frac{Q^2}{Q^2 +m_X^2 -m^2}\,,
\ee
from which we can verify that the elastic scattering case, $m = m_X$, corresponds to $x= 1$. The standard limit in inelastic scattering is the Bjorken limit, where $Q^2\to \infty$ and $\n \to \infty$, while $x$ is fixed. In this work, instead, we are interested in the regime of small $Q^2$, where non-perturbative effects are relevant. \\
The squared amplitude describing the lepton-nucleon scattering (\ref{scattproc}) can be written in terms of the leptonic ($l^{\mu \nu}$) and the hadronic tensor ($W^{\mu \nu}$) as (see e.g. \cite{Jaffe:1996zw, Deur:2018roz, Manohar:1992tz})
\be
\label{sqrtampl}
\bigg|\frac{\cal A}{4\pi}\bigg|^2 = \frac{\alpha_{em}^2}{Q^4} l^{\mu \nu} W_{\mu \nu}\,,
\ee
where $\alpha_{em}=e^2/4\pi^2\sim 1/137$ is the electromagnetic fine structure constant.  If on one hand, the leptonic tensor is completely known, on the other hand, the hadronic one, containing the whole hadronic information of the process, is hard to extract since it depends on non-perturbative QCD effects. In principle, it is given in terms of certain matrix elements of the hadronic electromagnetic current $J^{\mu}$ between the initial and final hadronic state(s) as
\be
W^{\mu\nu} = \frac{1}{4\pi} \sum_{X} \langle p, s| J^{\mu}| X \rangle \langle X | J^{\nu}| p, s \rangle (2\pi)^4 \delta(p+q-p_X)\,,
\label{formula1}
\ee
or, equivalently, as \footnote{In (\ref{formula2}) the subscript $_c$ underlines that we are considering the connected graphs.}
\be
W^{\mu\nu} = \frac{1}{4\pi} \int d^4 x e^{iq\cdot x} \langle p, s| [J^{\mu}(x), J^{\nu}(0)] | p,s\rangle_c\,,
\label{formula2}
\ee
where $p$ and $s$ are the momentum and spin of the target nucleon. Moreover, here we follow the conventions of \cite{Jaffe:1996zw} where states are chosen to be covariantly normalized as
\be
\label{norm}
\langle p|p'\rangle = 2 \sqrt{E E'} (2\pi)^3 \delta^3 (p-p')\,.
\ee
$W^{\mu\nu}$ is commonly expressed also in terms of the so-called \textit{unpolarized} and \textit{polarized} structure functions, respectively $F_1(x, Q^2)$, $F_{2}(x, Q^2)$ and  $g_1(x,Q^2)$, $g_2(x,Q^2)$: these are scalar dimensionless functions which encode all the information on the internal structure of the nucleon. Indeed, from symmetries constraints, we can parameterize the symmetric $(W^{\{\mu\nu\}})$  and anti-symmetric $(W^{[\mu\nu]})$ parts of the hadronic tensor as  \footnote{We are following the notation in \cite{Bayona:2011xj}, where a mostly plus signature of the metric is chosen. In a mostly minus signature case, as in \cite{Jaffe:1996zw}, the terms in equation (\ref{wsim}) and (\ref{wantisim}) are multiplied by a $-1$ factor.  Indeed, it is easy to check that the two definitions are mutually consistent.} \footnote{Here, we are focusing on a spin $1/2$ target, as the nucleon, for which the symmetric part of the hadronic tensor (\ref{wsim}) is independent of the nucleon spin $S$, and the spin-dependent part of $W_{\m\n}$ is antisymmetric in $\m,\n$. Notice that this is not true for higher spin targets.}
\be
\label{wsim}
W^{\{\mu\nu\}} =\left(\eta^{\mu\nu} - \frac{q^{\mu} q^{\nu}}{q^2} \right) F_1 -\left[ \left(p^{\mu} - \frac{\nu}{q^2} q^{\mu}\right) \left(p^{\nu} - \frac{\nu}{q^2} q^{\nu}\right)\right] \frac{F_2}{\nu}\,,
\ee
and
\be
\label{wantisim}
W^{[\mu\nu]} = i \epsilon^{\mu\nu\lambda\sigma} q_{\lambda} \left[ \frac{S_{\sigma}}{\nu} (g_1+g_2) - \frac{q\cdot S}{\nu^2} p_{\sigma}\, g_2\right]\,.
\ee
Here the nucleon spin polarization (pseudo)vector $S^{\mu}$, which is orthogonal to the target momentum $p\cdot S =0$, is normalized as $
S^2 =  m^2$.
Notice that current conservation implies that the hadronic and leptonic tensors satisfy the conditions 
\be
q_\m l^{\m\n} = 0\,, \hspace{1cm}  q_\m W^{\m\n} =  q_\m \big[W^{\{\mu\nu\}}+W^{[\mu\nu]} \big]= 0\,.
\ee 
Studying the full inclusive inelastic scattering process and computing the structure functions at low energy is an extremely daunting task. In the present work we will just focus on contributions arising from the production of single final state baryons with spin $1/2$ and $3/2$. Hence, the route we will follow (within the holographic QCD approach) consists in computing the hadronic tensor $W^{\mu\nu}$ at low $Q^2$ (so to truncate the infinite sum over hadronic states $X$) using its expression as written in eq. (\ref{formula1}). For each contribution, we will thus have to compute the matrix element of the electromagnetic current between the final state $X$ and the initial nucleon. 
\subsection{The Breit frame}
Let us consider an exclusive $\gamma\, B\rightarrow B_X$ transition process. In the rest frame of the target baryon $B$, the 3-momenta of the (virtual) photon and that of the outgoing hadron $B_X$ are obviously collinear. This is a particular example of a Breit frame, where the photon and hadrons 3-momenta all lie on the same axis. Working in such kind of frames is helpful, for instance, when adding the helicities of the incoming and outgoing particles. A particular choice we will use in the following is $E= E_X$, which means that the photon has zero energy $q^0 =0$. Setting the momentum of the photon to be negative and along the $x^3$ direction, the baryon and photon momenta in the Breit frame can be written as 
\ba
\label{breit1}
&&q^\m = (0,0,0, -Q),\hspace{1cm} Q= \sqrt{Q^2},\nb\\ 
&&p^\m = (E,0,0, \ms p),\hspace{1cm} \ms p = \frac{Q}{2x},\nb\\
\label{breit3}
&& p_X^\m = q^\m + p^\m.
\ea
For an exclusive $\gamma\, B\rightarrow B_X$ transition process, in the above frame  $p$ and $p_X$ are both positive if $Q^2< m_X^2-m^2$. This is the regime we will be mostly focus on.\footnote{Typically, instead, the Breit frame is chosen in such a way that the 3-momenta of the incoming and outgoing hadron have opposite directions. See \cite{Carlson:2003je} for relevant observations on the matter.} 
\subsection{The helicity amplitudes}
\label{sec:helicity}
The baryonic resonance ($B_X$) contribution to the nucleon structure functions is encoded in the helicity amplitudes associated to $\g \, B\to B_X$ transitions \cite{Carlson:1998gf} (see also \cite{Ramalho:2019ocp,Aznauryan:2008us}), where $B=N$ (nucleon). These are  
in turn related to matrix elements of the baryon electromagnetic current projected along the photon polarization vectors.

According to the conventions in \cite{Carlson:1998gf}, the production process of a single sharp baryonic resonance in the Breit frame defined in the previous subsection, is described by the following matrix elements\footnote{Notice that in \cite{Ramalho:2023hqd}, the nucleon and the final resonance states are dimensionless, while in \cite{Carlson:1998gf} the electromagnetic current matrix element has a dimension of a mass, so that the prefactor $1/(2m)$ is introduced to make amplitudes dimensionless. In what follows, in analyzing observables such as spin polarizabilities, we will choose the convention in \cite{Carlson:1998gf} for the helicity amplitudes.}
\ba
\label{Gi}
G^+_{B_XB} &=&\frac{1}{2m} \big\langle B_X, h_X = + 1/2| \e^{+}_\m J^\m| B, h = -1/2 \big \rangle\,,	\nb\\
G^-_{B_XB} &= &\frac{1}{2m} \big\langle B_X, h_X = + 3/2| \e^{+}_\m J^\m| B, h =  1/2 \big \rangle\,,	\nb\\
G^0_{B_XB} &=& \frac{1}{2m} \big\langle B_X, h_X=  1/2| \e^{0}_\m J^\m| B, h =  1/2 \big \rangle\,,
\ea
where $h, h_X$ are the initial and final helicities, and 
\be
\e^{\mu}_{\pm} = \frac{1}{\sqrt{2}}(0,\mp 1, -i,0)\,,\hspace{1cm} \e^{\mu}_{0} = \frac{1}{Q} (\ms q,0,0, q^0)\,,
\label{polvecgen}
\ee
are the polarization vectors of the virtual photon with momentum $q^\m = (q^0,0,0, \ms q)$, which, in the Breit frame defined in the previous subsection has to be set to $q^\m = (0,0,0, - Q)$.

Notice that if the final baryon has spin 1/2, $G_{B_XB}^{-}$ must be absent.

Contracting quadratic combinations of the photon polarization vectors with the hadronic tensor $W_{\mu\nu}$ in (\ref{formula1}) it is possible to realize that each sharp baryonic resonance gives the following contributions to the nucleon structure functions \cite{Carlson:1998gf} 
\ba
\label{carlson}
F_{1}(x,Q^2) &=& \sum_{m_X} \d((p+q)^{2} +m_{X}^{2})m^{2} \bigg(| G^{+}_{B_XB}|^2+| G^{-}_{B_XB}|^2\bigg),\nb\\
\bigg(1+\frac{\nu^{2} }{Q^{2}m^{2}}\bigg)F_{2}(x,Q^2) &=& -\sum_{m_X} \d((p+q)^{2} +m_{X}^{2})\n \bigg[ | G^{+}_{B_XB}|^2+|G^{-}_{B_XB}|^2  + 2| G^{0}_{B_XB}|^2\bigg],\nb\\
\bigg(1+\frac{Q^{2}m^{2}}{\nu^{2} }\bigg)g_{1}(x,Q^2) &=& \sum_{m_X} \d((p+q)^{2} +m_{X}^{2})m^{2} \bigg[(| G^{+}_{B_XB}|^2-| G^{-}_{B_XB}|^2 +\nb\\
&& -(-1)^{S_X- 1/2}\frac{Qm\sqrt{2}}{\n} G^{+\, *}_{B_XB} G^{0}_{B_XB}\bigg],\nb\\
\bigg(1+\frac{Q^{2}m^{2}}{\nu^{2} }\bigg)g_{2}(x,Q^2) &=& -\sum_{m_X} \d((p+q)^{2} +m_{X}^{2})m^{2} \bigg[| G^{+}_{B_XB}|^2-| G^{-}_{B_XB}|^2 +\nb\\
&&+ (-1)^{S_X- 1/2}\eta_X\frac{\n\sqrt{2}}{Qm}G^{+\, *}_{B_XB} G^{0}_{B_XB} \bigg].
\ea
Here, $S_X$ and $\eta_{X}$ are the spin and the parity of the resonance. 

In order to adapt the previous relations to the realistic case of non-sharp resonances, one can replace the Dirac delta functions with approximated expressions involving the resonances' decay widths. We will consider this approximation in section \ref{sec:spinpol}.
\subsubsection*{Elastic scattering}
Although we are not interested in elastic scattering processes, it is useful to look at the behavior of the above-introduced quantities in the elastic case $m_X =m$. In this limit, we get
\be
x\to 1\,, \hspace{1cm} \n \to -\frac{Q^2}{2}\,.
\ee
For elastic (Born) scattering, the helicity amplitudes are related to the electric ($G_E$) and magnetic ($G_M$) Sachs form factor by \cite{Carlson:1998gf}
\be
G_+(Q^2) \equiv G^+_{BB}(Q^2) = \frac{Q}{m\sqrt {2}} G_M(Q^2)\,,\hspace{1cm} G_0(Q^2) \equiv G^0_{BB}(Q^2) = G_E(Q^2)\,,
\ee
and the polarized structure functions in this limit read (see e.g. \cite{Drechsel:2002ar})
 \bea
 (g_1+g_2)_{Born} &=& \frac{G_E(Q^2) G_M(Q^2)}{2} \delta(1-x)\,, \\
 \left(g_1 - \frac{4 m^2}{Q^2} g_2\right)_{Born}& =& \frac12 G^2_M(Q^2) \delta(1-x)\,.
 \eea
Sachs form factors are respectively normalized at $Q^2=0$ as $
G_E(0) = e_N$ and $G_M(0) = \m_N$, 
where $e_N$ and $\m_N$ are the charge (in units of $e$) and the total magnetic moment (in units of $e/2m$) of the respective nucleon.  Then, we can see $G^0_{B_XB} $ and $G^+_{B_XB} $ as objects somehow carrying information about the nucleon electric and magnetic distributions. 
\subsection{Spin polarizabilities}
The generalized spin polarizabilities describe the collective spin-dependent response of the nucleon to an external electromagnetic field. They can be expressed as moments of certain combinations of the spin structure functions. In particular, the longitudinal-transverse and the forward spin polarizabilities are given respectively by
\be
\label{expdeltaLT}
\delta_{LT}(Q^2) = \frac{16\alpha_{em} m^2}{Q^6} \int_{0}^{x_0} dx\, x^2\left[g_1(x,Q^2) + g_2(x,Q^2)\right]\,,
\ee
and 
\be
\label{expgamma0}
\gamma_0(Q^2) = \frac{16\alpha_{em} m^2}{Q^6} \int_{0}^{x_0} dx\, x^2\left[g_1(x,Q^2) -\frac{4 m^2 x^2}{Q^2} g_2(x,Q^2)\right]\,,
\ee
where $x_0$ is the so-called pion-production threshold
\be
x_0 = \frac{Q^2}{Q^2+(m_{\pi} +m)^2-  m^2}\,,
\label{ics0}
\ee
which excludes the elastic scattering contribution from the integrals (with $m_{\pi}$ being the pion mass).
We refer to e.g. \cite{Jaffe:1996zw, Deur:2018roz} for more details about the derivation and properties of spin polarizabilities. We just notice that, due to the $x^2$ weighting in the integrands, both observables are expected to be dominated by the region close to the pion-production threshold.
\subsubsection{Resonance contributions to $\d _{LT}$}
Using (\ref{carlson}), we can obtain the (sharp) baryonic resonance contributions to $\d _{LT}$ in (\ref{expdeltaLT}). Let us rewrite the sum  of $g_1(x,Q^2)$ and $g_2(x,Q^2)$ as 
\be
\label{g1g2tot}
g_1(x,Q^2) + g_2(x,Q^2) =-(-1)^{S_X- 1/2}\eta_X  \sum_{m_X} \d((p+q)^{2} +m_{X}^{2})m\sqrt 2 \frac \n Q G^{+\, *}_{B_XB} G^{0}_{B_XB}\,.
\ee
For each resonance with mass $m_X$, it is possible to rewrite
\begin{multline}
\label{dtildex}
\d((p+q)^{2}+m_{X}^{2}) = \d\bigg(1-x\big( 1+ \frac{(m_X^2-m^2)}{Q^2}\big)\bigg) \left| \frac{1}{2p\cdot q}\right|
= \d(\tilde x_X-x) \frac{\tilde x^2_X}{ Q^2}\,,
\end{multline}
where
\be
 \tilde x_X = \frac{Q^2}{ Q^2 + m_X^2-m^2}\,.
\ee
Evaluating the integral over $x$ in (\ref{expdeltaLT}), each sharp $X$-resonance will contribute to $\d_{LT}$ only if $\tilde x_X < x_0$, namely if
\be
 \frac{Q^2}{Q^2 +m_{X}^2 -m^2}< \frac{Q^2}{Q^2+(m_{\pi}+ m)^2 -m^2}\,. 
\ee
This request is satisfied by the baryonic resonances since they all have masses $m_X>m + m_{\pi}$. In light of these considerations, from (\ref{expdeltaLT}) we see that, in the limit of sharp resonances, their contribution to the longitudinal-transverse spin polarizability takes the form
\ba
\label{dlt3}
\delta_{LT}(Q^2) &=& \frac{16\alpha_{em} m^2}{Q^6} \sum_{m_{X}} (-1)^{S_X- 1/2}\eta_X \tilde x^3_X\frac{m\sqrt 2 }{2Q} G^{+\, *}_{B_XB} G^{0}_{B_XB}=\nb\\
&=&  \frac{8\sqrt 2 \alpha_{em} m^3}{Q} \sum_{m_{X}} \frac{(-1)^{S_X- 1/2}\eta_X}{ (Q^2+m_X^2-m^2)^3} G^{+\, *}_{B_XB}G^{0}_{B_XB}\,.
\ea
\subsubsection{Resonance contributions to $\g_0$}
The forward spin polarizability is expressed as an integral over a combination of the spin structure functions as shown in (\ref{expgamma0}). 
Using eq. \eqref{carlson} we can express the resonance contribution to that combination as
\be
\label{combg1g2g0}
 g_1(x,Q^2) -\frac{4m^2 x^2}{Q^2} g_2(x,Q^2) = \sum_{m_{X}} \d((p+q)^{2}+m_{X}^{2})m^2  \bigg[| G^{+}_{B_XB}|^2-| G^{-}_{B_XB}|^2\bigg]\,.
\ee
Using again (\ref{dtildex}) and the considerations made in the previous subsection, the sharp-resonance contribution to the forward spin polarizability turns out to be given by
\be
\label{god2}
\gamma_0(Q^2)
=  16\alpha_{em} m^4\sum_{m_{X}} \frac{1}{  (Q^2+m_X^2-m^2)^4} \bigg[| G^{+}_{B_XB}|^2-| G^{-}_{B_XB}|^2\bigg]\,.
\ee
Notice how the $ G^{-}_{B_XB}$ helicity amplitude, which is non-zero only for processes involving helicity $3/2$ resonances, gives a negative contribution to the forward polarizability.  
\section{The WSS Holographic QCD model}
\label{holorev}
As we have anticipated in the Introduction, our aim is to compute the resonance contribution to the nucleon spin structure functions and hence to the generalized spin polarizabilities, using the holographic gauge/gravity correspondence within the Witten-Sakai-Sugimoto (WSS) model \cite{Witten:1998zw,Sakai:2004cn}. The WSS model is a $3+1$ dimensional QCD-like $SU(N_c)$ theory with $N_f$ quarks and further adjoint matter fields, arising from a Kaluza-Klein reduction from higher dimensions, whose mass scale is denoted by $M_{KK}$. The effective holographic description of QCD-like hadrons in the WSS model with $N_f\ll N_c$ massless quarks\footnote{We defer considering the inclusion of quark masses, along the lines of \cite{Aharony:2008an, Hashimoto:2008sr}, to a future work.} is provided by a five-dimensional $U(N_f)$ Yang-Mills-Chern-Simons theory with action  
\begin{equation}
\label{ymcs}
S_f = -\kappa\int d^4x d z\, \left(\frac{1}{2}h(z)  \, \Tr\mathcal{F}_{\mu\nu}\mathcal{F}^{\mu\nu} +   k(z)  \Tr\mathcal{F}_{\mu z}\mathcal{F}^\mu_{\;\; z}\right)
 + \frac{N_c}{24\pi^2}\int  \omega_5(\mathcal{A}) \,,
\end{equation}
where $z\in(-\infty,\infty)$ is the holographic direction, four-dimensional indices $\mu,\nu$ are raised by the flat Minkowski metric and (in $M_{KK}=1$ units)
\be
\kappa = \frac{N_c\lambda}{216\pi^3}\,,\quad h(z) = (1+z^2)^{-1/3}\,,\quad k(z) = (1+z^2)\,.
\ee 
Moreover
\begin{equation}
\omega_5(\mathcal{A}) =  \Tr\left(\mathcal{A}\wedge \mathcal{F}^{ 2}- \frac{i}{2}\mathcal{A}^{ 3} \wedge\mathcal{F} - \frac{1}{10}\mathcal{A}^{ 5}\right)\;,\quad d \omega_5(\mathcal{A}) = \Tr\mathcal{F}^{  3}\,.
\label{omega5}
\end{equation}
The holographic description is valid in the planar $N_c\gg1$ limit, at strong coupling $\lambda\gg1$. Here $\lambda$ is a 't Hooft-like parameter proportional to the ratio of the Yang-Mills string tension and $M_{KK}^2$. 

It is customary to decompose the $U(N_f)$ gauge field $\mc A$ into $U(1)$ and $SU(N_f)$ parts
\be
\mc A = A + \frac{1}{\sqrt{2N_f}}\hat A = A^a T^a + \frac{1}{\sqrt{2N_f}}\hat A\,,
\ee
where $T^a \, (a=1,.., N_f^2 -1)$ are the generators for $SU(N_f)$\footnote{ $T^a$ are normalized as $\Tr(T^a T^b) = \frac{1}{2} \d^{ab}$.}. In terms of this decomposition the action (\ref{ymcs}) can be written as
\ba
\label{action5d}
S_{f} &=& -\frac{\k}{2} \int{d^4xdz}\,\bigg( \frac{1}{2}h(z)\hat F^{\m\n} \hat  F_{\m\n} + k(z) \hat F^{\m z} \hat F_{\m z}\bigg) + \nb\\
&&-\k \int{d^4xdz}\, \Tr  \bigg[\frac{1}{2}h(z) F^{\m\n} F_{\m\n} + k(z)  F^{\m z} F_{\m z}\bigg]+ \nb\\
&&+ \frac{N_c}{24\pi^2} \int{ \bigg(\omega^{SU(N_f)}_5( A)} + \frac{3}{\sqrt{2N_f}} \hat A \w \Tr F^2 +  \frac{1}{2\sqrt{2N_f}} \hat A \w \hat F^2\bigg)\,.
\ea
Mesons in the model correspond to gauge field fluctuations. For instance, in the simplest $N_f=1$ case, their effective description arises considering the following expansions
\bea
{\mathcal A}_{\mu}(x^{\mu},z) &=& \sum_{n=1}^{\infty} B_{\mu}^{(n)}(x^{\mu})\psi_n(z)\,,\nonumber \\
{\mathcal A}_{z} (x^{\mu},z) &=& \sum_{n=0}^{\infty} \varphi^{(n)}(x^{\mu})\phi_n(z)\,.
\eea
The functions $\psi_n(z)$, $\phi_n(z)$ form complete sets and are normalized in such a way that the fields $B_{\mu}^{(n)}$ and $\varphi^{(n)}$ get canonical kinetic and mass terms in four dimensions. Each function $\psi_n(z)$ satisfies the eigenvalue equation
\be \label{eqforpsi}
-h(z)^{-1} \partial_z (k(z)\partial_z\psi_n(z))=\lambda_n \psi_n(z)\,,
\ee
with normalization condition $\kappa\int dz h(z) \psi_n(z) \psi_m(z) =\delta_{mn}$. Moreover
\be
\phi_n(z) = \lambda_n^{-1/2}\partial_z\psi_n(z) \quad (n>1)\,,\quad \phi_0(z) = (\kappa\pi)^{-1/2}k(z)^{-1}\,.
\ee 
The modes $B_{\mu}^{(n)}$ correspond to massive axial (for even $n$) and vector (for odd $n$) fields with masses $m_n^2 = \lambda_n M_{KK}^2$.  For instance $B_{\mu}^{(1)}$ and $B_{\mu}^{(2)}$ correspond, respectively, to the $\rho$ and the $a_1$ meson. The mode $\varphi^{(0)}$ corresponds to the pion, while $\varphi^{(n)}$ with $n\ge1$ are eaten by the $B_{\mu}^{(n)}$. 

Remarkably, after reducing the action (\ref{action5d}) to four dimensions, one gets not only the chiral Lagrangian, with pion decay constant  $f_{\pi}^2 = 4\kappa M_{KK}^2/\pi$ and the Skyrme term, but also the effective action and coupling for the other massive meson fields. 

For $N_f\ge2$, baryons in the WSS model are described by instanton-like solutions of the gauge field satisfying
\be
\label{nbclass}
n_B = \frac{1}{64\pi^2}\int{d^3xdz \, \e_{MNPQ} F^a_{MN} F^a_{PQ}}\,,
\ee
where the instanton number $n_B$ coincides with the baryon number.\footnote{See \cite{sheet} for a discussion on WSS baryons in the $N_f=1$ case, based on \cite{komar}.}

In this work we will take $N_f =2$ and consider $n_B=1$ baryons. The classical instanton solution can be found numerically following the results in \cite{Hashimoto:2008zw}. It can be easily expressed around $z=0$ \cite{Hata:2007mb} where it reads
\bea \label{instanton}
&&A_M^{\mathrm{cl}} = -i (1-f(\xi)) g^{-1} \partial_M g\,, \qquad A_0^{\mathrm{cl}}=\widehat{A}_M=0\,,\nonumber \\
&&\widehat{A}^{\mathrm{cl}}_0 = \frac{N_c}{8\pi^2\kappa}\frac{1}{\xi^2}\left[1-\frac{\rho^4}{(\rho^2+\xi^2)^2}\right]\,,
\eea
where $M=1,2,3,z$, $\xi^2 \equiv (z-Z)^2 + |\vec{x}-\vec{X}|^2$ and 
\begin{equation}
f(\xi) = \frac{\xi^2}{\xi^2+\rho^2}\,, \quad g(x) = \frac{(z-Z)\mathbf 1 - i (\vec{x}-\vec{X})\cdot \vec{\tau}}{\xi}\,,
\end{equation}
where $\vec{\tau}$ are the Pauli matrices.

The solution displays a set of parameters: the instanton center of mass position $X^{M}$ in the 4d Euclidean space, the instanton size $\rho$, and three $SU(2)$ angular parameters accounting for the fact that the solution can be rotated by a global gauge transformation.  The latter are usually expressed as
\be
\label{modulia}
 {\bf{a}}= a_4 \mathbf 1 + i a_a \t^a, \hspace{1cm} (a=1,2,3),
\ee
with the condition $a_1^2+a_2^2+a_3^2+a_4^2 =1$. 

The on-shell value of the action (\ref{action5d}) on the above solution, is minimized for $Z = Z_{cl}=0$, $\rho^2 =\rho_{cl}^2= (N_c/8\pi^2 \kappa)\sqrt{6/5}$. The remaining parameters are instead genuine instanton moduli. Without loss of generality, below we will take $\vec X =0$. Moreover from the on-shell value of the action one gets the classical baryon mass scaling as $m\sim \lambda N M_{KK}$.

Baryons are thus heavy objects and their quantum description is provided by a non-relativistic Hamiltonian written in terms of the above (pseudo) moduli promoted to time-dependent operators. Baryonic states are then the eigenstates of this Hamiltonian. Their wave-functions are of the form\footnote{Here, at this level, the baryonic states are normalized to 1, while in the following we will move to a different normalization, according to the conventions in \cite{Carlson:1998gf}, i.e. $\le B, \vec p\,^\pri, s' |B, \vec p,s\re = 2m (2\pi)^3 \d_{s^\pri s} \d^3(\vec p\,^\pri-\vec p)$.}
\be
\vert \vec p, B, s\rangle = \vert \vec p\rangle \vert B,s\rangle =e^{i\vec p \cdot \vec X} \vert B,s\rangle\,.
\ee
They are characterized by $\vec p, s$, the latter being the eigenvalue of the third component of the spin operator, and by the quantum numbers $B=(l,I_3,n_{\rho},n_z)$. Here, $l/2$ (with $l$ integer) gives the spin $S$ and isospin $I$ of the state (all the WSS baryons have $I=S$), $I_3$ is the eigenvalue of the third component of the isospin and $n_{\rho},n_{z}$ are quantum numbers related to the operators $\rho$ and $Z$.

For instance, the $l= 1$ states correspond to $I=S=1/2$ baryons, which include the nucleons, and the $l=3$ ones correspond to $I=S=3/2$ states, which include the $\D$. Heavier baryons with a common spin and isospin can be represented by states with non-vanishing values of $n_\r$ and $n_z$. Furthermore, excited states with an odd $n_z$ correspond to odd parity baryons. A detailed review on the explicit expression for the baryonic states we will need in our analysis is given in appendix \ref{appa}.

The baryon mass spectrum (in units $M_{KK}=1$ and modulo a missing subtraction of the zero-point vacuum energy \cite{Hata:2007mb}) is given by
\be
\label{massformula}
M = M_0 + \sqrt{\frac{(l+1)^2}{6} +\frac{2}{15}N_c^2} + \frac{2(n_\r +n_z) +2}{\sqrt{6}}\,,
\ee
where $M_0 = 8\pi^2 \k$. This formula reproduces the expected large $N_c$ behavior $M\sim{\cal O}(N_c)$ and indicates that low-lying resonances are nearly degenerate with nucleons in the strict holographic limit.\footnote{See \cite{Imaanpur:2022lvp} for a discussion of possible subleading corrections to the above mass formula in the strong coupling regime.}

As usual in the large $N_c$ limit, WSS baryons are heavy and nearly stable objects. In most of this work we will thus treat the excited baryonic states as strictly sharp resonances.
\subsection{Electromagnetic current}
For $N_f=2$ WSS baryons, the electromagnetic current, a combination of the isoscalar $\hat J_V^{\m} $ and the isovector $J_V^{3\m}$ components, is holographically given by \cite{Hashimoto:2008zw}
\begin{equation}
\label{current}
J_{\mu}=J_{V\, \m}^{3}+ \frac{1}{N_c} \hat J_{V\, \m}= -\kappa\left[k(z)\Tr(F_{\mu z}\tau^3)+\frac{k(z)}{N_c}{\widehat F}_{\mu z}\right]^{z\to\infty}_{z\to-\infty}\,,
\end{equation}
where the field strength $F$ corresponds to the classical instanton solution describing the baryons. At the quantum level the current is promoted to an operator acting on the baryonic states. The components of the vector currents operators take the form \cite{Hashimoto:2008zw, Bayona:2011xj}
\ba
\label{curhatjo2}
\hat J^0_{V} &=& \frac{N_c}{2} G_{V},\nb\\
 \hat J^{i}_{V} &=&-\frac{N_c}{2M_0} \big[\p_i H_V P_Z -G_VP^i -\frac{S_a}{2}\big((\p_i\p_a-\d^{ia}\p_j^2)H_{V}+\e^{ija}\p_jG_{V}\big)\big],\nb\\
J^{c,0}_{V} &=&\frac{M_0}{4} \big\{ \r^2\Tr[\t^c\p_0({\bf{a}}\t^a{\bf{a}}^{-1})\big]\p_a H_{V} +\frac{4I^c}{M_0} G_{V}\nb\\
&& - 	\r^2 \Tr[\t^c{\bf{a}}\t^a{\bf{a}}^{-1}]\dot X^i \big((\p_i\p_a-\d^{ia}\p_j^2)H_{V}-\e^{iaj}\p_jG_{V}\big)\big\},\nb\\
\label{curji2}
J_{V}^{c,i}&=& -\frac{M_0}{4}\r^2  \Tr[\t^c{\bf{a}}\t^a {\bf{a}}^{-1}]\big((\p_i\p_a-\d^{ia}\p_j^2)H_{V}+\e^{ija}\p_jG_{V}\big).
\ea
Here $P^i$ and $P_Z$ are the canonical momenta conjugated to $X^i$ and $Z$ respectively and $I^a$ and $S^a$ are the components of the isospin and spin operators (see eq. (\ref{IS}) for their explicit expression). Moreover
\ba
&&G_V(Z,r) = -\sum_{n=1}^\infty g_{v^n} \psi_{2n-1}(Z)Y_{2n-1}(r)\,,\nb\\
&&H_V(Z,r) = -\sum_{n=1}^\infty \frac{g_{v^n}}{\l_{2n-1}} \p_Z\psi_{2n-1}(Z)Y_{2n-1}(r)\,,
\ea
where $g_{v^n}$ are the decay constants of the vector mesons
\be
g_{v^n} = -2\k \big[k(z) \p_z \psi_{2n-1}(z)\big]_{z= +\infty} = m_{v^n}^2\k \int{dz \, h(z) \psi_{2n-1}}\,,
\ee
and $Y_n$ is the Yukawa potential with meson mass $m_n = \sqrt{\l_n}$ ($\l_0 = 0$),
\be
\label{yuk}
Y_n(r) = -\frac{1}{4\pi}\frac{e^{-\sqrt{\l_n}r}} {r}\,, \hspace{1cm} r=|\vec x-\vec X|\,.
\ee
The matrix elements of the Fourier transformed isoscalar and isovector currents read \cite{Bayona:2011xj}
\ba
\label{BXJB}
\langle p_X,B_X, h_X| \hat J_V^{0}(0)|p,B,h\rangle &=&\frac{N_c}{2} \langle h_X, I_3^X|h,I_3\rangle F^1_{B_XB}(\vec q\,^2)\d_{n_{\r X}n_\r}\,,\nb\\
\langle p_X,B_X, h_X| \hat J_V^i(0)|p,B,h\rangle &=& \frac{N_c}{4M_0} \, \langle h_X, I_3^X| \bigg\{F^1_{B_XB}(\vec q\,^2) \big[2p^i- i \e^{ija}q_jS_a\big]\nb\\
&&+ 2q^i F^3_{B_XB}(\vec q\,^2)- F^2_{B_XB}(\vec q\,^2) (q^{i}q^{a}- \vec{q}\,^{2}\d^{ia})S_a\big]\bigg\}
|h,I_3\rangle \d_{n_{\r X}n_\r}\,,\nb\\
\langle p_X,B_X, h_X| J_V^{c,0}(0)|p,B,h\rangle &=& \frac{1}{4} \langle h_X, I_3^X|\langle n_{\r X}| \nb\\
&&\times\bigg\{F^1_{B_XtB}(\vec q\,^2) \big[4I^c +i \epsilon^{ija}p^i q_{j}\rho^{2} \Tr  [ \tau^{c}\textbf{a}\tau^{a}\textbf{a}^{-1}]\big] \nb\\
&&+ F^2_{B_XB}(\vec q\,^2)\bigg[-M_0i q_{a}  \rho^{2} \Tr  [ \tau^{c}\partial_{0}(\textbf{a}\tau^{a}\textbf{a}^{-1})] \nb\\
&&+ (\vec{p}\cdot\vec{q} q_{a} -\vec{q}^2p_{a})\rho^{2} \Tr  [ \tau^{c}\textbf{a}\tau^{a}\textbf{a}^{-1}]\bigg]\bigg\}|n_\r\rangle |h,I_3\rangle\,,\nb\\
\langle p_X,B_X, h_X| J_V^{c,i}(0)|p,B,h\rangle &=& \frac{M_{0}}{4}\big[i F^1_{B_XB}(\vec q\,^2) \epsilon^{ija}q_{j}   + F^2_{B_XB}(\vec q\,^2)(q^{i}q^{a} - \vec{q}\,^{2}\delta^{ia})\big]\nb\\
&& \times \langle n_{\r X}|\rho^{2}|n_\r\rangle\, \langle h_X, I_3^X|\ \Tr [ \tau^{c}\textbf{a}\tau^{a}\textbf{a}^{-1}] |h,I_3\rangle\,,
\ea
where 
\ba
\label{CD}
F^1_{B_XB}(\vec q\,^2)  &=& \sum_{n =1}^{\infty}\frac{ g_{v^{n}}\le \psi_{2n-1}(Z) \re}{\vec{q}^{2} + \lambda_{2n-1}}\,, \nb\\
F^2_{B_XB}(\vec q\,^2) &=&  \sum_{n =1}^{\infty}\frac{ g_{v^{n}}\le \partial_{Z}\psi_{2n-1}(Z) \re}{\lambda_{2n-1}(\vec{q}^{2} + \lambda_{2n-1})}\,.
\ea
\section{WSS helicity amplitudes}
\label{sec:helicity}
Since, to leading order in the holographic regime, the WSS baryons are quantum eigenstates of a non-relativistic Hamiltonian, it will be useful to work in a Breit frame where the time component of the photon momentum is set to zero. 
Thus, in analyzing the helicity amplitudes in the model we will refer to expressions in (\ref{Gi}), working in the Breit frame (\ref{breit1}), with the corresponding polarization vectors (\ref{polvecgen}) given as
\be
\label{polvec}
\e_{\pm\,\m} = \frac{1}{\sqrt{2}}(0,\mp 1, -i,0)\,,\hspace{1cm} \e_{0\,\mu} = (1,0,0,0)\,.
\ee
Moreover we will use the results for the matrix elements of the electromagnetic current collected in (\ref{BXJB}). Let us thus consider each amplitude in detail.
\subsection{Scalar amplitude $G^0_{B_XB} $}
\label{subsecG0}
With the above conventions, we have
\ba
\label{G0}
G^0_{B_XB}  &=& \frac{1}{2m} \big\langle B_X, h_X  = + 1/2| J^0| B, h = + 1/2 \big \rangle =\nb\\
&=&\frac{1}{2m}  \big \le \frac{1}2 \big(1+2 I^3\big) \d_{\eta_X,\eta} F^{1}_{B_XB}(\vec q\,^{2}) -i  \d_{\eta_X,-\eta}\frac{M_0}4 F^2_{B_XB}(\vec q\,^2) \ms q \, \rho^{2} \Tr  [ \tau^{3}\partial_{0}(\textbf{a}\tau^{3}\textbf{a}^{-1})] \re.\nb\\
\ea
Moreover, it is useful to recall the following identity \cite{Ballon-Bayona:2012txi}
\be
\Tr(\t^3\p_0( {\bf{a}}\t^3{\bf{a}}^{-1})) = -\frac{2i}{M_0\r^2}\bigg( a_4\frac{\p}{\p a_4} - a_1\frac{\p}{\p a_1}-a_2\frac{\p}{\p a_2}+ a_3\frac{\p}{\p a_3}\bigg)\,,
\ee
from which it is easy to see that 
\ba
&&\Tr(\t^3\p_0( {\bf{a}}\t^3{\bf{a}}^{-1})) | n , h  \big \rangle = -2h  \frac{2i}{M_0\r^2} | n , h \big \rangle,\nb\\
&& \Tr(\t^3\p_0( {\bf{a}}\t^3{\bf{a}}^{-1})) | p , h \big \rangle = +2 h\frac{2i}{M_0\r^2} | p , h \big \rangle.
\ea
Thus we can list all the possible results\footnote{Here $\tilde B_X$ collect all baryon quantum numbers except $n_{z X}$ and $n_{\rho X}$. Analogously for $\tilde n$ and $\tilde p$.}:
\begin{itemize}
\item{\bf{Neutron-Positive parity resonance}}: $G^0_{B_XB}  = 0$ ;
\item{\bf{Neutron-Negative parity resonance}}: $G^0_{B_XB}  = - \frac{1}{2}F^2_{B_XB}(\vec q\,^2)\, \ms q  \frac{1}{2m}\le \tilde B_X, 1/2| \tilde n, 1/2\re$ ;
\item{\bf{Proton-Positive parity resonance}}: $G^0_{B_XB}  =  F^{1}_{B_XB}(\vec q\,^{2})\frac{1}{2m}\d_{n_{\r X}n_\r} \le \tilde B_X, 1/2| \tilde p, 1/2\re$ ;
\item {\bf{Proton-Negative parity resonance}}: $G^0_{B_XB}  = + \frac{1}{2}F^2_{B_XB}(\vec q\,^2)\, \ms q \frac{1}{2m}\le \tilde B_X,  1/2| \tilde p, 1/2\re $ .
\end{itemize}
Notice that if the final resonance belongs to an isospin representation with $I > 1/2$ (as the $\D$ resonances), $G^0_{B_XB} $ vanishes identically. This is in some sense expected since the $\g N \to \D$ transition overwhelmingly involves transverse photons; moreover, this qualitatively agrees with the experimental observation that, in this case, the scalar amplitude $G^0_{B_XB}$ is suppressed with respect to $G^+_{B_XB}$ and $G^-_{B_XB}$.

\subsection{Longitudinal amplitude $G^+_{B_XB}$}
In the Breit frame (\ref{breit1}, with the polarization vectors (\ref{polvec}), the expression for $G^+_{B_XB}$ in (\ref{Gi}) reduces to 
\begin{multline}
\label{G+}
G^+_{B_XB} =-\frac{1}{2\sqrt 2m} \big\langle B_X,  1/2| (J^1 +iJ^2)| B, -1/2 \big \rangle = \\
=- \frac{1}{4 \sqrt 2 M_0}\frac{1}{2m} \big\le  \d_{\eta_X,\eta} F^{1}_{B_XB}(\vec q\,^{2}) \ms q \big[ \big(S_1+ iS_2\big)- M_0^2  \r^2  \big(O_1 +iO_2\big)\big]+\\ + \d_{\eta_X,-\eta}F^{2}_{B_XB}(\vec q\,^{2}) \ms q^2  \big[ \big(S_1 +iS_2\big)-M_0^2  \r^2  \big(O_1 +iO_2\big)\big]\big\re=\\
= - \frac{1}{4 \sqrt 2 M_0}\frac{1}{2m} \big\le (  \d_{\eta_X,\eta}F^{1}_{B_XB}(\vec q\,^{2}) \ms q  + \d_{\eta_X,-\eta}F^{2}_{B_XB}(\vec q\,^{2}) \ms q^2 )\big[ \big(S_1+ iS_2\big)- M_0^2  \r^2  \big(O_1 +iO_2\big)\big]\re,
\end{multline}
where we have defined the (non-diagonal isospin) operator as
\be
\label{Oa}
O^a =  \Tr [ \tau^{3}\textbf{a}\tau^{a}\textbf{a}^{-1}]\,.
\ee
Furthermore, notice that 
\be
\big(S_1+i S_2\big) | n\, (p) , -1/2  \big \re = S_+ | n\, (p) , -1/2  \big \re =    | n\, (p) , +1/2  \big \re\,.
\ee
Then  (\ref{G+}) reads
\begin{multline}
\label{G+1}
G^+_{B_XB} =- \frac{1}{4 \sqrt 2M_0} i ( \d_{\eta_X,\eta}F^{1}_{B_XB}(\vec q\,^{2}) \ms q  + \d_{\eta_X,-\eta}F^{2}_{B_XB}(\vec q\,^{2}) \ms q^2 )\times\\\times\frac{1}{2m} \bigg[-i \d_{n_{\r X}n_\r} \le \tilde B_X, 1/2| \tilde n (\tilde p), 1/2\re-  M_0^2\big\le  \r^2  \big(O_2 -iO_1\big)\re\bigg]=\\
=- \frac{1}{4 \sqrt 2M_0} ( \d_{\eta_X,\eta}F^{1}_{B_XB}(\vec q\,^{2}) \ms q  + \d_{\eta_X,-\eta}F^{2}_{B_XB}(\vec q\,^{2}) \ms q^2 ) \frac{1}{2m}\bigg[ 2\sqrt{m_Xm}\,\d_{I_X,I} \d_{n_{\r X}n_\r}-i M_0^2\le  \r^2 \re \le\big(O_2 -iO_1\big)\re\bigg].
\end{multline}

\subsection{Longitudinal amplitude $G^-_{B_XB}$}
The longitudinal amplitude $G^-_{B_XB}$ is different from zero only when the final resonance has a spin greater than 1/2; in Breit frame it reads
\begin{multline}
G^-_{B_XB} =-\frac{1}{2\sqrt 2 m} \big\langle B_X,  3/2| (J^1 +iJ^2)| B, 1/2 \big \rangle= \\
=  -\frac{1}{4 \sqrt 2 M_0}\frac{1}{2m} \big\le ( \d_{\eta_X,\eta}F^{1}_{B_XB}(\vec q\,^{2}) \ms q  + \d_{\eta_X,-\eta}F^{2}_{B_XB}(\vec q\,^{2}) \ms q^2 )\big[ \big(S_1+ iS_2\big)- M_0^2  \r^2  \big(O_1 +iO_2\big)\big]\re\,,
\end{multline}
so that
\ba
\label{G-}
G^-_{B_XB} 
&=&  \frac{M_0}{4 \sqrt 2}\frac{1}{2m}\big\le ( \d_{\eta_X,\eta}F^{1}_{B_XB}(\vec q\,^{2}) \ms q  + \d_{\eta_X,-\eta}F^{2}_{B_XB}(\vec q\,^{2}) \ms q^2 )  \r^2  \big(O_1+iO_2\big)\re.
\ea
\section{Extrapolation to QCD and comparison with experimental data}
\label{sec:compare}
According to eq. (\ref{massformula}) the resonance-nucleon mass difference in the WSS model is given by
\be
m_X - m = M_{KK}\left[\sqrt{\frac{(l+1)^2}{6}+\frac{2N_c^2}{15}} -\sqrt{\frac{2}{3}+\frac{2N_c^2}{15}}\right] + M_{KK}\frac{2(n_{\r _X} +n_{z_ X})}{\sqrt{6}}\,,
\ee
as the ground state nucleon, of mass $m$, has quantum numbers $l=1\,,n_{\rho}=n_{z}=0$. Since we are interested in comparing model predictions with QCD data for nucleon observables in the first resonance regions, we will set, as in \cite{Hata:2007mb} 
\be
\label{barparam1}
N_c=3\,,\quad M_{KK}\approx 488\, \text{MeV}\,,
\ee 
so to reproduce the difference between the measured mass of the $\Delta(1232)$ resonance (corresponding to the $l=3\,, n_{\rho}=n_{z}=0$ WSS state) and the average nucleon mass\footnote{Since we work in the chiral limit, the WSS nucleons have the same mass. See \cite{Bigazzi:2018cpg} for a study where the introduction of isospin-breaking mass terms allows to remove this degeneracy.} $m\approx 939\, \text{MeV}$. In addition, we will fix $\lambda$ as in \cite{Fujii:2022yqh}, requiring that 
\be
\label{barparam2}
M_0 = m \approx 939\, \text{MeV}\,,\quad{\rm so\,that}\,\, \l \approx 54.4\,. 
\ee
Let us recall that in the literature it is also common to fit the WSS model parameters with mesonic data\footnote{In detail, setting the $\rho$ meson mass $m_{\rho} = \sqrt{\lambda_1} M_{KK}$ and the pion decay constant $f_{\pi}^2 = 4\kappa M_{KK}^2/\pi$ to their experimental values.} setting
\be
N_c=3\,, \hspace{1cm}M_{KK} \approx 949\, \text{MeV}, \hspace{1cm} \l \approx 16.6\,. 
\label{standardpar}
\ee
With this choice, however, the baryon mass spectrum sensibly departs from the experimental one \cite{Hata:2007mb} and thus, fixing the parameters as in (\ref{standardpar}) would produce a systematic error that we want to avoid.\footnote{In fact, using (\ref{standardpar}), we have carried out the same analysis as the one in the following sections, finding that the corresponding spin polarizabilities are suppressed by about one order of magnitude compared to those derived from the choices (\ref{barparam1}), (\ref{barparam2}). Our analysis also suggests that the results are more sensitive to variations of $M_{KK}$ than of $\lambda$.}

As we have shown in the previous section, the helicity amplitudes for $\gamma\,B\rightarrow B_X$ transitions, with $B=N$ and $B_X$ a spin $1/2$ or spin $3/2$ baryon resonance, depend on the transition form factors  $F^{1}_{B_{X}B}(Q^{2} )$, $F^{2}_{B_{X}B}(Q^{2} )$ defined in (\ref{CD}).  The former can be written as \footnote{In a generic frame. In Breit frame $Q^2 = \vec q\, ^2$.}
\be
\label{f1num}
F^{1}_{B_{X}B}(Q^{2} ) = \sum_n\frac{g_{v^n}g_{v^nB_XB}} {Q^2 + \l_{2n-1}}\,,
\ee
where
\be
\label{gv1}
g_{v^nB_XB} \equiv \langle n_{z_X}| \psi_{2n-1}(Z)|n_z\rangle\,,
\ee
are the effective couplings between a vector meson, a positive parity baryon $B_X$ (i.e. $n_{z_X}$ is an even integer; when $n_{z_X}$ is odd, i.e. the resonance has negative parity, the above coupling is zero) and the baryon $B$. In a similar way, $F^{2}_{B_{X}B}(Q^{2} )$ is expressed as
\be
\label{f2num}
F^{2}_{B_{X}B}(Q^{2} ) = \sum_n\frac{g_{v^n}\tilde g_{v^nB_XB}} {Q^2 + \l_{2n-1}}\,,
\ee
where
\be
\tilde g_{v^nB_XB} \equiv \frac{1}{\l_{2n-1}} \langle n_{z_X}| \partial_Z\psi_{2n-1}(Z)|n_z\rangle\,,
\label{gv2}
\ee
which is different from zero only if $n_{z_X}$ is an odd integer. The above expressions highlight the emergence of vector-meson dominance in the model. 

The eigenfunctions $\psi_{2n-1}(z)$ and the corresponding eigenvalues can be found numerically solving the differential equations (\ref{eqforpsi}) as done in \cite{Sakai:2004cn}. The first eigenvalues are listed in Table \ref{table:eigen}. Using these results, and the expressions for the baryon eigenfunctions given in appendix \ref{appa}, we can numerically evaluate the expressions in (\ref{gv1}) and (\ref{gv2}) to get the effective couplings $g_{v^nB_XB}$ and $\tilde g_{v^nB_XB} $ for the first baryon states (in our analysis, those with $n_{z_X}=0,1,2$). 
\begin{table}[h!]
\centering
\begin{tabular}{|c| c| c| c|c|c|c|} 
 \hline
    \rowcolor{lightgray} $n$ & 1& 2 & 3&4&5 &6\\ [0.5ex] 
 \hline
 $\l_{2n-1}$ &0.69 & 3.16 &7.66& 14.31&23.12& 34.13\\[1ex] 
 \hline
\end{tabular}
\caption{\small Some numerical values for the mesonic eigenvalues.}
\label{table:eigen}
\end{table}
Then, we can obtain approximate expressions for $F^{1}_{B_{X}B}(Q^{2} )$ and $F^{2}_{B_{X}B}(Q^{2} )$. Since we are interested in the low-$Q^2$ regime, the infinite sums over the tower of vector meson states in (\ref{f1num}) and (\ref{f2num}), can be approximated by including just the first few vector mesons states in the numerical computations (actually, in our computation, we have accounted for the first 32 meson states).
\subsection{Helicity amplitudes}
Commonly, experimental results on nucleon-resonance transitions\footnote{Actually, most of the available data refer to proton-resonance transitions.} are expressed in terms of the helicity amplitudes $A^{1/2}_{B_XB}$, $A^{3/2}_{B_XB}$ and $S^{1/2}_{B_XB}$ which are defined in terms of the amplitudes in (\ref{Gi}) as follows \cite{Ramalho:2023hqd,Blin:2021twt}\footnote{In order to compare our results with data  collected in \cite{Ramalho:2023hqd}, we will follow here its convention for the amplitudes $G^{\pm,0}_{B_XB}$ which differ from the ones in \cite{Carlson:1998gf} by a factor of $\sqrt{\frac{m_X}m}$.}:
\ba
\label{defhelamp}
&&A^{1/2}_{B_XB}  = \sqrt{\frac{2\pi\a_{em}}{K}}G^+_{B_XB}\,,\nb\\
&&S^{1/2}_{B_XB}  = \eta_X(-1)^{S_X-1/2}\sqrt{\frac{2\pi\a_{em}}{K}}\frac{|\ms q_R|}{Q}G^0_{B_XB}\,,\nb\\
&&A^{3/2}_{B_XB}  =\eta_X(-1)^{S_X-1/2}\sqrt{\frac{2\pi\a_{em}}{K}}G^-_{B_XB}\,,
\ea
where $K = \frac{m_X^2-m^2}{2m_X}$ and
\be
|\ms q_R| = \frac{\sqrt{\big((m_X+m)^2 +Q^2\big)\big((m_X-m)^2 +Q^2\big)}}{2m_X}\,,
\ee
is the spatial momentum of the photon in the resonance rest frame.

Having in mind a comparison with experimental data, it is important to recall that, just as in the Skyrme model, the WSS baryonic states arise as eigenstates of a non-relativistic Hamiltonian. We can thus ask what happens if we give them a more appropriate relativistic structure as it has been worked out in \cite{Bayona:2011xj,Ballon-Bayona:2012txi, Boschi-Filho:2011jen} for the spin $1/2$ case. There, disregarding possible relativistic corrections to the non-relativistic effective Lagrangian of the collective modes, spin $1/2$ baryonic states have been taken to have the spinor-like structure\footnote{Here we are using the spinor normalization of \cite{Ramalho:2023hqd}.}
\be
\label{Relspinor}
u_B(p,h)= \sqrt{ \frac{E+m}{2m}} \bpm 1 \\ \frac{ \vec p \cdot \vec \s }{E +m} \epm  |B, h\re\,,
\ee
where $ |B, h\re $ are the non-relativistic WSS baryonic states. Thus, nucleon-resonance matrix elements in (\ref{G0}) and (\ref{G+}) could be replaced  (in Breit frame) by
\be
\le B_X, h_X| \mc O|B,h \re_{\text{Rel.}} = \mc F_\pm \le B_X, h_X| \mc O|B,\l \re\,,
\label{relshift}
\ee
where we have defined
\be
\mc F_\pm = \sqrt{ \frac{(E+m)(E + m_X)}{4m m_X}}\bigg(1\mp \frac{\ms p_X \ms p}{(E+m)(E + m_X)}\bigg)\,.
\label{fpm}
\ee
In what follows, the results derived using this approximate relativistic formula for the spin $1/2$ states will be shown in figures by purple lines.
 
For the WSS spin 3/2 resonances similar relativistic shifts have not been computed yet\footnote{Technically, in the WSS model we do not have direct access to a Rarita-Schwinger-like object describing the final resonance states; essentially the main difficulties are due to the missing relation between the “spin 1” part of the Rarita-Schwinger field and the $SU(2)$ space moduli coordinates. We leave this interesting issue to future works.}: we will thus take the same relations as in (\ref{relshift}) and (\ref{fpm}) as possible indications.

\subsubsection{$N(1440)$ resonance}
\begin{figure}[htb]
\begin{center}
\includegraphics*[width=0.45\textwidth]{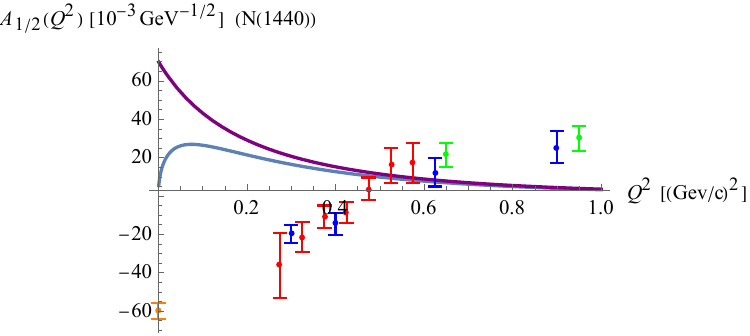}
\end{center}
\caption{\small
$A^{1/2}_{B_XB}(Q^2)$ 
helicity amplitude in unit of $ 10^{-3} \text{GeV}^2$ for the $\g \,p \to N(1440)$ transition. The solid blue line represents our results, in the non-relativistic limit, divided by a factor of 3. The solid purple line gives the result (divided by 3) from the approximate relativistic approach (\ref{Relspinor}).
The experimental data are from CLAS analysis of $\pi^+ \pi^- p$ electroproduction \cite{Mokeev2012,Mokeev2015} (red  and green points) and $N\pi$ electroproduction off protons \cite{CLAS:2009ces} (blue points).   }
\label{HAp1440}
\end{figure}

\begin{figure}[htb]
\begin{center}
\includegraphics*[width=0.45\textwidth]{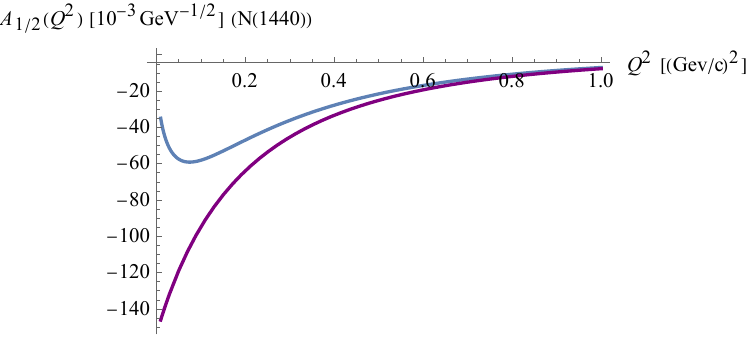}
\end{center}
\caption{\small $A^{1/2}_{B_XB}(Q^2)$ 
helicity amplitude for the $\g\, n \to N(1440)$ transition in the WSS model. The solid blue (resp. purple) line corresponds to the non-relativistic (resp. approximate relativistic) result. }
\label{HAn1440}
\end{figure}

In the WSS model, the spin $J^P = 1/2^+$ Roper resonance N(1440) corresponds to the $(l=1\,, n_{\rho}=1\,,n_z=0)$ state (see appendix \ref{appa}). According to the results collected in section \ref{subsecG0}, being a positive parity resonance it gives rise, to leading order in the holographic limit, to a vanishing scalar amplitude $G^0_{B_XB}$ in the case of the transition involving the neutron. In the same limit, $G^0_{B_XB}$ is also vanishing for the proton case, due to the presence of the $\d_{n_{\r X},n_\r}$ factor in the amplitude.  On the other hand, $G^+_{B_XB}$ is nonzero for both the neutron and proton transitions. In figure \ref{HAp1440}, we show our results for the $\g \, p\to N(1440)$ $A^{1/2}_{B_XB}(Q^2)$ helicity amplitude in comparison with experimental data from CLAS Collaboration \cite{CLAS:2009tyz, CLAS:2009ces, Mokeev2012 ,Mokeev2015}. The $A^{1/2}_{B_XB}(Q^2)$ helicity amplitude for the $\g \, n\to N(1440)$ transition is given in figure \ref{HAn1440}, while $S^{1/2}_{B_XB}(Q^2)$ is zero for both the nucleons due to the vanishing $G^0_{B_XB}$ (\ref{G0}).\footnote{This leading order result is in contradiction with the experimental data on $S^{1/2}_{B_XB}(Q^2)$ for the proton-Roper transition. Here one can argue that, possibly, the WSS result will be modified by subleading $1/\lambda$ or $1/N_c$ effects; another possibility, explored in \cite{Fujii:2022yqh}, is that a different non-gauge-invariant definition of currents proposed in \cite{Hata:2008xc}, might cure the mismatch. In this case a much better agreement with the behavior of $A^{1/2}_{B_XB}(Q^2)$ for the proton-Roper transition is found too.}

At least from the analysis of the amplitudes involving the proton, where data at $Q^2>0$ are available, we can conclude that the WSS model, at this stage, does not provide reliable predictions on the Roper transitions and, then, on Roper's contribution to the nucleon spin polarizabilities.

\subsubsection{$N(1535)$ resonance}
The next spin 1/2 resonance we consider is the negative parity N(1535) resonance ($J^P = 1/2^-$). It corresponds (see appendix \ref{appa}) to the $(l=1\,, n_{\rho}=0\,,n_z=1)$ WSS state. In this case, the scalar amplitude $G^0_{B_XB}$  (\ref{G0}) is non-vanishing for both neutron and proton; thus we will have nonzero contributions to the longitudinal-transverse spin polarizability for both nucleons. 

\begin{figure}[htb]
\begin{center}
\includegraphics*[width=0.45\textwidth]{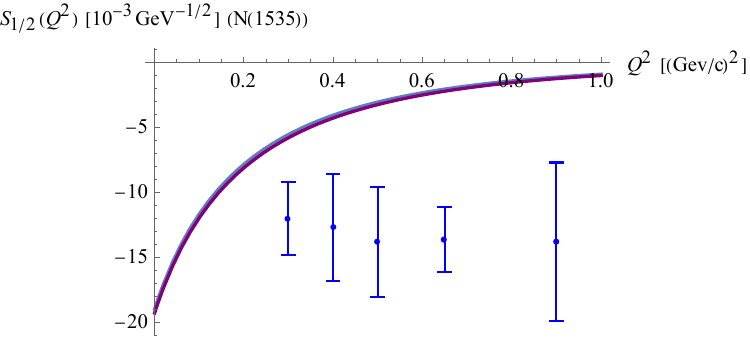}
\includegraphics*[width=0.45\textwidth]{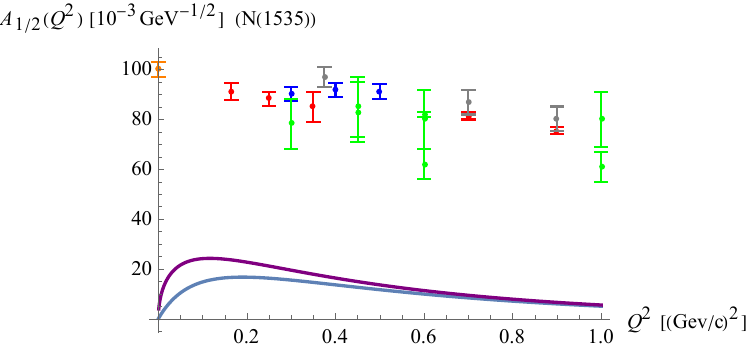}
\end{center}
\caption{\small $S^{1/2}_{B_XB}(Q^2)$ (left) and $A^{1/2}_{B_XB}(Q^2)$ (right)
helicity amplitudes for the $\g \,p \to N(1535)$ transition. The solid blue (resp. purple) lines correspond to the non-relativistic (resp. ``relativistic'') result. The CLAS data are from the analysis of $N\pi$ photo-production off proton \cite{CLAS:2009tyz} (red points), electroproduction off proton \cite{CLAS:2009ces}  (green points) and $\eta N$ electroproduction \cite{CLAS:2007bvs, CLAS:2000mbw}  (blue and gray points).}
\label{HAp1535}
\end{figure}

Our findings for the proton are shown in figure \ref{HAp1535}, where experimental data from CLAS Collaboration \cite{CLAS:2009tyz,CLAS:2009ces, Burkert:2002zz,CLAS:2007bvs,CLAS:2000mbw} are displayed too. In this case we can notice a better qualitative agreement between WSS prediction and real world data. 

\begin{figure}[htb]
\begin{center}
\includegraphics*[width=0.45\textwidth]{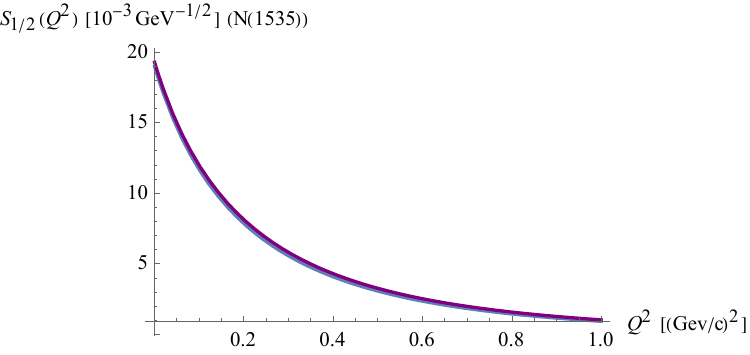}
\includegraphics*[width=0.45\textwidth]{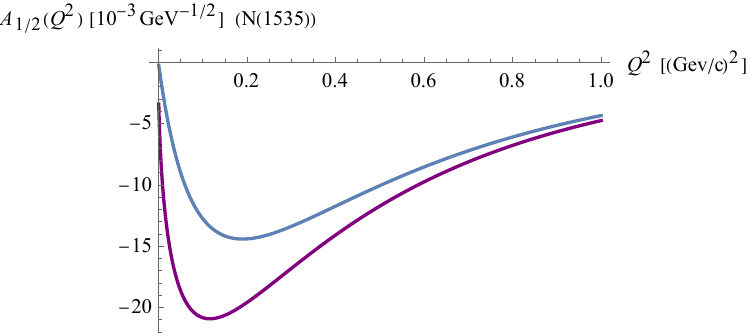}
\end{center}
\caption{\small $S^{1/2}_{B_XB}(Q^2)$ and $A^{1/2}_{B_XB}(Q^2)$ 
helicity amplitudes for the $\g \,n \to N(1535)$ transition. The solid blue (resp. purple) lines correspond to the non-relativistic (resp. ``relativistic'') result.}
\label{HAn1535}
\end{figure}

The neutron results (for which experimental data {\bf at $Q^2>0$} are not available at the moment) are shown in figure \ref{HAn1535}.

\subsubsection{$\D(1232)$ resonance}
Finally, let us consider the lightest nucleon resonance, the spin 3/2 $\D(1232)$ resonance. In the WSS model its charge eigenstates are identified with baryon states with quantum numbers $l= 3$ and $n_{z} = n_{\r} = 0$ (see appendix \ref{appa}). Being the first of the low-lying nucleon resonances it is generically expected to sensibly contribute to the low-$Q^2$ behavior of the nucleon structure functions. 

The $\Delta$ helicity amplitude does not depend on the isospin of the target nucleon. This can be easily realized first observing that the isoscalar current does not contribute the $\g N \to \D$ transition, since it cannot induce a transition from a spin $1/2$ state to a spin $3/2$ one
\be
\langle 3/2, I_{3,\D}| J^\m|1/2, I_3\rangle  =  (1/2, I_3, 1,0| 3/2, I_{3,\D})\langle 3/2|| J^{\t=3, \m}||1/2\rangle\,,
\ee
since $(1/2, I_3, 0,0| 3/2, I_{3,\D}) = 0$. Moreover, the Clebsch-Gordan coefficient $ (1/2, I_3, 1,0| 3/2, I_{3,\D})$ is equal to $\sqrt{2/3}$ for both the $\g\, n\to \D^0$ and $\g\, p\to \D^+$ transitions, implying equal contribution of the transition to the structure functions of neutron and proton.

In the WSS model, to leading order in the holographic regime, for any spin $3/2$ resonance, the scalar amplitude $G^0_{B_XB}$  (\ref{G0}) (and so the longitudinal helicity amplitude $S^{1/2}_{B_XB}(Q^2)$) vanishes identically. In fact, just as it happens in the Skyrme model to leading order in the large $N_c$ limit (see e.g. \cite{Braaten:1986iw}), there is no (quadrupole-like) operator in the time component of the electromagnetic current which can allow for an isospin transition between the nucleon (isospin representation $I=1/2$) and $\D$ states (isospin representation $I=3/2$).\footnote{This observation holds in a generic frame and not just in the Breit one; in fact, thanks to the conservation of the current, one can reduce  $G^0_{B_XB}(Q^2)$  to the evaluation of the matrix element of the temporal component of the electromagnetic current.} 
This leading order result qualitatively agrees with the experimental observation that, for the $\Delta$ states, $S^{1/2}_{B_XB}(Q^2)$ is actually suppressed with respect to  $A^{1/2}_{B_XB}(Q^2)$ and $A^{3/2}_{B_XB}(Q^2)$.

\begin{figure}[htb]
\begin{center}
\includegraphics*[width=0.45\textwidth]{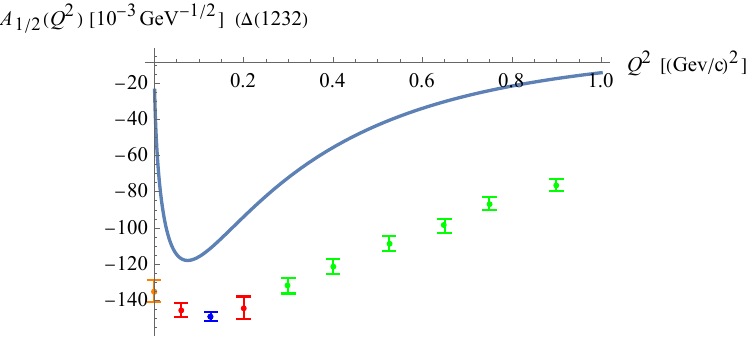}
\includegraphics*[width=0.45\textwidth]{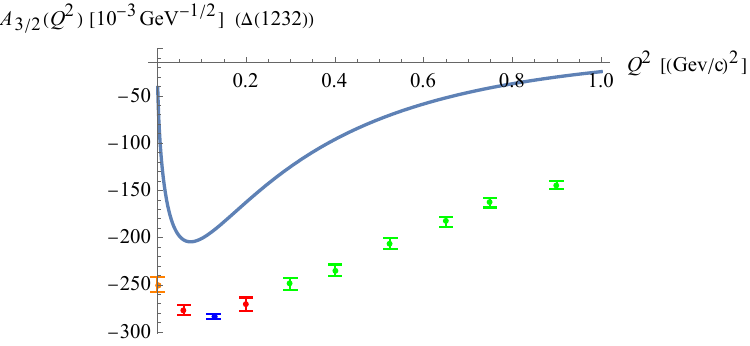}
\end{center}
\caption{\small $A^{1/2}_{B_XB}(Q^2)$ and $A^{3/2}_{B_XB}(Q^2)$ 
helicity amplitudes for the $\g \,p \to \D^+(1232)$ ($\g \,n \to \D^0(1232)$) transition. The solid blue line represents our results for the low-$Q^2$ behavior of $A^{1/2}_{B_XB}(Q^2)$ (left panel) and $A^{3/2}_{B_XB}(Q^2)$ (right panel). The experimental data are from MAMI  (red points), MIT/Bates   (blue points) and CLAS (green points) analysis of $N\pi$ electroproduction off protons \cite{A1:2008ocu, OOPS:2004kai,CLAS:2009ces}.}
\label{HApD}
\end{figure}

The results for the $A^{1/2}_{B_XB}(Q^2)$ and $A^{3/2}_{B_XB}(Q^2)$ helicity amplitudes for the $\g \,p \to \D^+(1232)$ transition are shown in figure \ref{HApD} with experimental data from \cite{A1:2008ocu, OOPS:2004kai,CLAS:2009ces}. According to the previous observations, the related results for the neutron-$\D^0(1232)$ transition are the same as in figure \ref{HApD}. From (\ref{G+1}), (\ref{G-}) and (\ref{o1o2}), we can easily observe that in our approach $A^{3/2}_{B_XB}(Q^2) = \sqrt 3 A^{1/2}_{B_XB}(Q^2)$ to leading order in the holographic limit. This leads to a vanishing electric amplitude $E$ for the $\D(1232)$, since $E =A^{1/2}_{B_XB}(Q^2)- A^{3/2}_{B_XB}(Q^2)/\sqrt 3$ \cite{Ramalho:2023hqd}.  \footnote{This is consistent with the vanishing amplitude $S^{1/2}_{B_XB}(Q^2)$, since the two quantities are related by the so-called Siegert's theorem \cite{Ramalho:2023hqd}.}

As we have already observed, for helicity amplitudes involving spin $3/2$ states we still miss the precise form of the relativistic shift introduced in the spin $1/2$ cases. Nevertheless, we expect that, just as for the above mentioned spin $1/2$ cases, the shift will affect the quantitative behavior of the helicity amplitudes, leaving the qualitative one essentially unaffected. 

\subsubsection*{Next low-lying resonances}
What we have found above concerning the helicity amplitudes for the first three excited WSS baryonic states, namely N(1440), N(1535) and $\D(1232)$ can be extended naturally, using expressions in Section \ref{sec:helicity} for the amplitudes $G^{\pm,0}_{B_X B}$, to the next low-lying resonances, e.g. $\D(1600)$ $(l=3\,, n_{\rho}=1\,,n_z=0)$, the negative parity N(1650) $(l=1\,, n_{\rho}=1\,,n_z=1)$ and N(1710) $(l=1\,, n_{\rho}=0\,,n_z=2)$. We do not report here the related results but we will consider the contributions of these resonances too in evaluating the low-$Q^2$ behavior of the spin polarizabilities.

\subsection{Single resonance contributions to $\d_{LT}(Q^2)$ }
Using the sharp-resonance approximation, and the expressions for the structure functions in terms of helicity amplitudes as in \cite{Carlson:1998gf}, the contribution of a single nucleon resonance to the longitudinal-transverse spin polarizability $\d_{LT}(Q^2)$ as can be extracted from (\ref{dlt3}) reads 
\ba
\label{rescontd}
\d_{LT}(Q^2)_X =(-1)^{(S_X-1/2)}\eta_X \frac{8\sqrt2 \a_{em} m^3}{Q} \frac{1}{(Q^2 + m_X^2 -m^2)^3}  G^{+ \, *}_{B_X B}G^0_{B_X B}.
\ea
In evaluating the above expression, following the prescription in \cite{Carlson:1998gf}, we have to normalize the nucleon and the resonance states to $2m$ and $2m_X$, respectively, using helicity amplitudes with an overall factor $(2m)^{-1}$ as in (\ref{Gi}).

For both nucleons, the contribution of the $\Delta(1232)$ (and higher spin $3/2$ resonances) to the longitudinal-transverse polarizability is negligible (in the WSS model, to leading order in the holographic limit, the contribution is actually zero, since $G^0_{B_XB}=0$ as we have previously observed). This feature agrees with what is observed in JLab experiments (see e.g. \cite{JeffersonLabE97-110:2019fsc}).  Moreover, the Roper contribution is vanishing for both nucleons too (once again since  $G^0_{B_XB}$ is zero in the model also for this transition).
\subsubsection{Neutron}
Since, in the neutron case, positive parity resonances give vanishing $G^0_{B_XB}$, the only nonzero contributions to  $\d_{LT}(Q^2)$ come from the (spin 1/2) negative parity resonances. Indeed, from (\ref{G0}) and (\ref{G+1}), we obtain
\begin{multline}
\d_{LT}(Q^2)_X =-\frac{ \a_{em} m^3}{M_0 Q} \frac{m_X}{m}\frac{1}{(Q^2 + m_X^2 -m^2)^3} (F^{2}_{B_XB}(\vec q\,^2))^2\, \ms q ^3\times\\ \times \big[ \d_{n_{\r X}n_\r}  -iM_0^2\le  \r^2 \re  \le\tilde  B_X, 1/2|\big(O_2 -iO_1\big)|\tilde n,-1/2\re\big]=\\
=-\text{sgn}(\ms q) \frac{m_X}{m} \frac{ \a_{em} m^3 Q^2}{M_0 (Q^2 + m_X^2 -m^2)^3} (F^{2}_{B_XB}(\vec q\,^2))^2 \big[ \d_{n_{\r X}n_\r}  -\frac 43 M_0^2\le  \r^2 \re\big],
\end{multline}
where we have used the WSS baryon states given in Appendix (\ref{appa}) and relations (\ref{o1o2}) for the matrix element of $O^a$ operators. 

\begin{figure}[htb]
\begin{center}
\includegraphics*[width=0.45\textwidth]{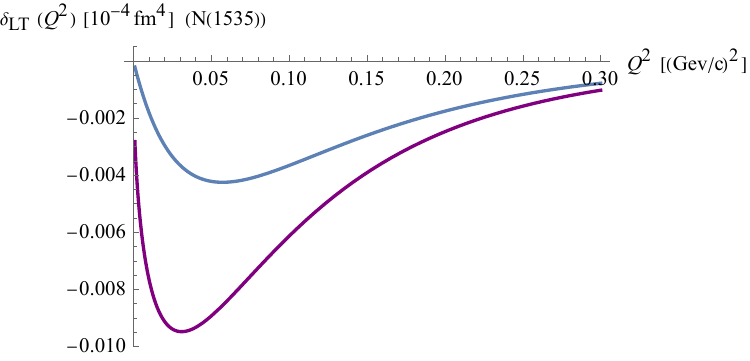}
\includegraphics*[width=0.45\textwidth]{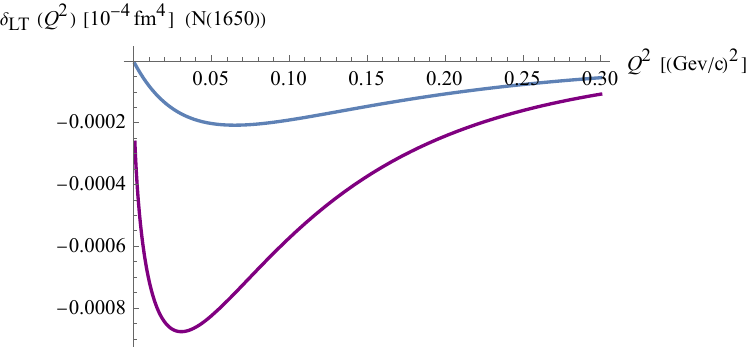}
\end{center}
\caption{\small Contributions to the neutron longitudinal-transverse spin polarizabilities from sharp N(1535) (left) and N(1650) (right) negative parity nucleon resonances in unit of $10^{-4} \text{fm}^4$. The solid blue (resp. purple) lines correspond to the non-relativistic (resp. ``relativistic'') results. }
\label{dn1535}
\end{figure}

In figure \ref{dn1535}, we present our results for the low-$Q^2$ behavior of the contribution of the $N(1535)$  and $N(1650)$ resonances to neutron's $\d_{LT}(Q^2)$. We see that $\d_{LT}(Q^2)$ is negative with increasing absolute value as $Q^2$ goes to zero.

\subsubsection{Proton}
In the proton case, spin $1/2$ positive parity resonances also contribute to $\delta_{LT}(Q^2)$, since the related scalar amplitude $G^0_{B_XB}$ is not vanishing. In what follows, we will make use again of (\ref{G0}), (\ref{G+}) and (\ref{o1o2}). 

Positive parity resonances give the following contributions
\begin{multline}
\d_{LT}(Q^2)_X =- \frac{2 \a_{em} m^3}{M_0 Q} \frac{m_X}{m}\frac{1}{(Q^2 + m_X^2 -m^2)^3} (F^{1}_{B_XB}(\vec q\,^{2}) )^2\ms q \times\\\times \big[\d_{n_{\r X}n_\r}  -i M_0^2\le  \r^2 \re  \le B_X, 1/2|\big(O_2 -iO_1\big)|p,-1/2\re\big]=\\
=- \text{sgn}(\ms q)  \frac{2 \a_{em} m^3}{M_0}  \frac{m_X}{m}\frac{1}{(Q^2 + m_X^2 -m^2)^3} (F^{1}_{B_XB}(\vec q\,^{2}) )^2\big[\d_{n_{\r X}n_\r}  +\frac 43 M_0^2\le  \r^2 \re\big]\,.
\end{multline}

\begin{figure}[htb]
\begin{center}
\includegraphics*[width=0.45\textwidth]{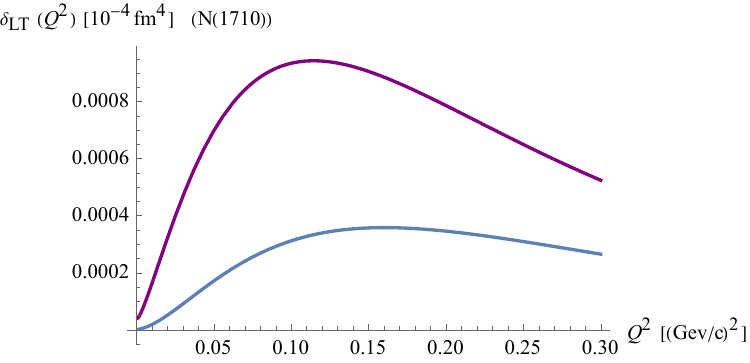}
\end{center}
\caption{\small Contribution to the proton longitudinal-transverse spin polarizability from the sharp positive parity nucleon resonance N(1710) in units of $10^{-4} \text{fm}^4$. The solid blue (resp. purple) line corresponds to the non-relativistic (resp. ``relativistic'') result.} 
\label{dp}
\end{figure}

\begin{figure}[htb]
\begin{center}
\includegraphics*[width=0.45\textwidth]{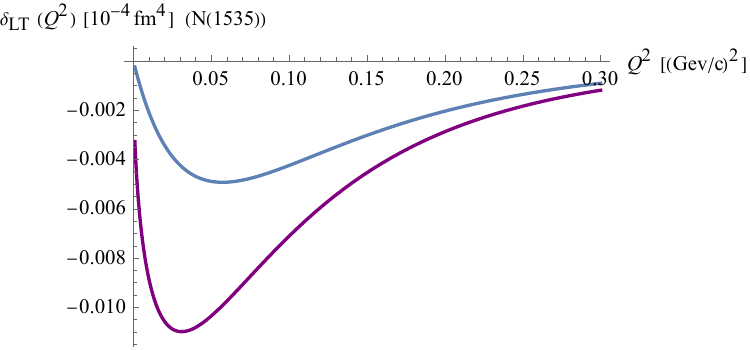}
\includegraphics*[width=0.45\textwidth]{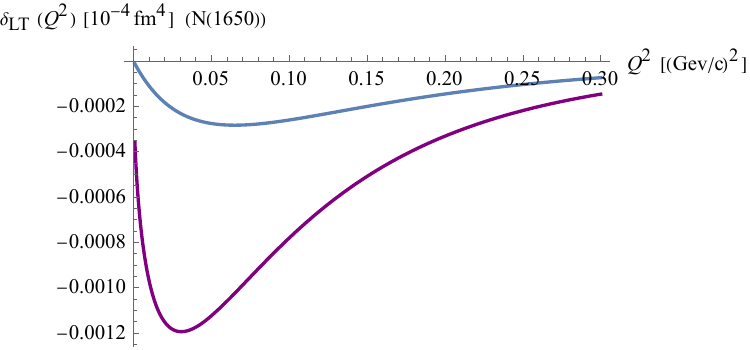}
\end{center}
\caption{\small Contributions to the proton longitudinal-transverse spin polarizability from sharp negative parity nucleon resonances N(1535) (left)  and N(1650) (right) in units of $10^{-4} \text{fm}^4$. The solid blue (resp. purple) lines correspond to the non-relativistic (resp. ``relativistic'') result.} 
\label{dp1}
\end{figure}

On the other hand, for negative parity resonances, we get
\begin{multline}
\d_{LT}(Q^2)_X = \frac{ \a_{em} m^3}{M_0 Q}  \frac{m_X}{m}\frac{1}{(Q^2 + m_X^2 -m^2)^3} (F^{2}_{B_XB}(\vec q\,^{2}) )^2\ms q^3 \times \\ \times \big[\d_{n_{\r X}n_\r}  -i  M_0^2\le  \r^2 \re \le B_X, 1/2|\big(O_2 -iO_1\big)|p,-1/2\re\big]=\\
=\text{sgn}(\ms q)  \frac{ \a_{em} m^3}{M_0}  \frac{m_X}{m}\frac{ Q^2}{(Q^2 + m_X^2 -m^2)^3} (F^{2}_{B_XB}(\vec q\,^{2}) )^2  \big[\d_{n_{\r X}n_\r}  +\frac 43 M_0^2\le  \r^2 \re\big]\,.
\end{multline}
In figures \ref{dp} and  \ref{dp1} we show the contributions of the first positive and negative parity resonances, i.e. N(1710) and N(1535), N(1650). In the former (resp. latter) case, we find a positive asymptotically decreasing (resp. negative increasing) function. 

\subsection{Single resonance contributions to $\g_{0}(Q^2)$ }
We can now analyze the single sharp resonance contributions to the forward spin polarizability $\g_{0}(Q^2)$ of the nucleons. From (\ref{god2}) it is given by
\be
\label{rescontg}
\g_0(Q^2)_X =  \frac{16 \a_{em} m^4 }{(Q^2 + m_X^2 -m^2)^4} \big[ |G^+_{B_XB}|^2- |G^-_{B_XB}|^2\big]\,.
\ee
Making use of (\ref{G+1}) and  (\ref{G-}) we can study separately the neutron and proton cases.
\subsubsection{Neutron}
In contrast with the longitudinal-transverse spin polarizability, $\g_{0}(Q^2)$ receives contributions from both spin 1/2 and spin 3/2 resonances of both parities.

Positive parity resonances give
\begin{multline}
\g_0(Q^2)_X =  \frac{16 \a_{em} m^4 }{(Q^2 + m_X^2 -m^2)^4}\frac{m_X}{m} \frac{1}{32 M_0^2}(F^{1}_{B_XB}(\vec q\,^{2}) )^2 \ms q^2\times \\\times\big[ |\d_{I_X,I} \d_{n_{\r X}n_\r}-M_0^2\le \r^2\re  \le \tilde B_X, 1/2| \big(O_1+iO_2\big)|\tilde n,-1/2\re|^2- |M_0^2\le \r^2\re\le \tilde B_X, 3/2| \big(O_1+iO_2\big)|\tilde n,1/2\re|^2\big]\\
=  \frac{ \a_{em} m^4 }{(Q^2 + m_X^2 -m^2)^4}\frac{m_X}{m}  \frac{1}{2 M_0^2}(F^{1}_{B_XB}(\vec q\,^{2}) )^2 Q^2 
\bigg[\d_{I_X,I}\bigg(\d_{n_{\r X}n_\r}- \frac43 M_0^2\le \r^2\re\bigg)^2-\d_{I_X,I+1}\frac{32} 9M_0^4 \le\r^2\re^2 \bigg].
\end{multline}

\begin{figure}[htb]
\begin{center}
\includegraphics*[width=0.45\textwidth]{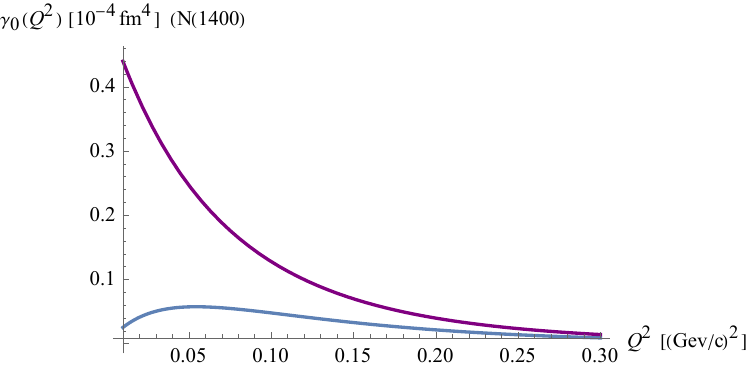}
\includegraphics*[width=0.45\textwidth]{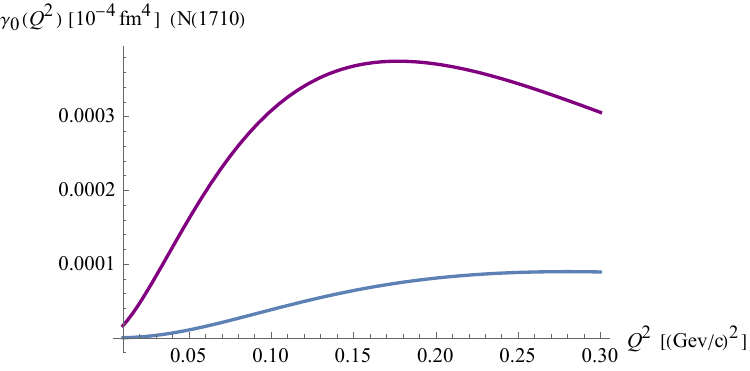}
\end{center}
\caption{\small Contributions to the neutron forward spin polarizability from sharp nucleon resonances N(1440) (left)  and N(1710) (right). The solid blue (resp. purple) lines correspond to the non-relativistic (resp. ``relativistic'') result.} 
\label{gn}
\end{figure}

\begin{figure}[htb]
\begin{center}
\includegraphics*[width=0.45\textwidth]{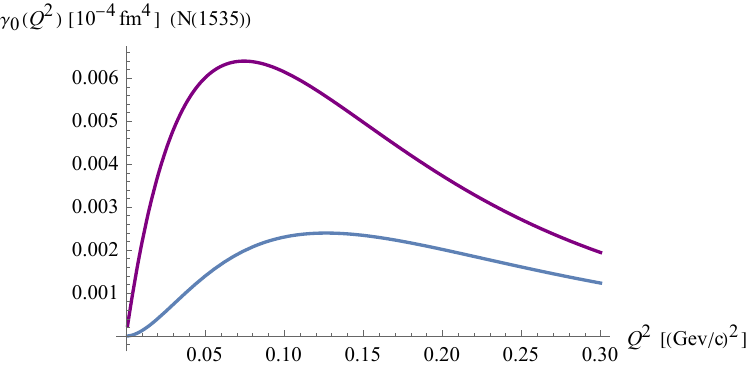}
\includegraphics*[width=0.45\textwidth]{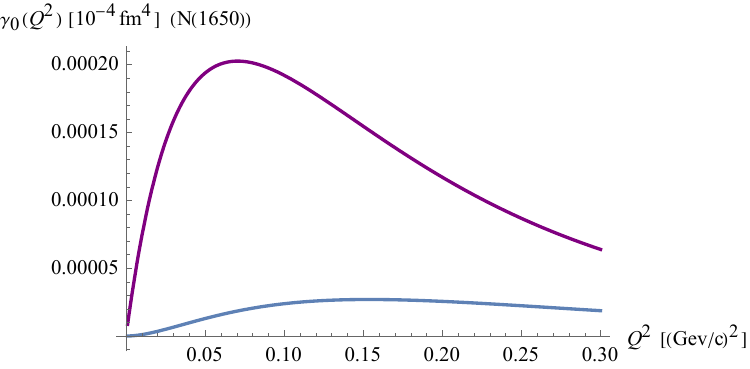}
\end{center}
\caption{\small Contributions to the neutron forward spin polarizability from sharp negative parity nucleon resonances N(1535) (left)  and N(1650) (right). The solid blue (resp. purple) lines correspond to the non-relativistic (resp. ``relativistic'') result.} 
\label{gn1}
\end{figure}

\begin{figure}[htb]
\begin{center}
\includegraphics*[width=0.45\textwidth]{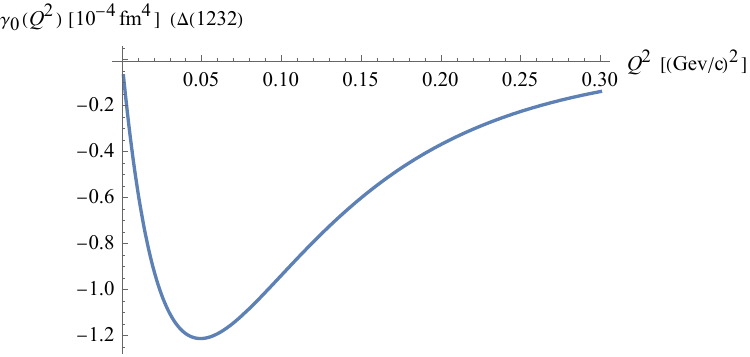}
\includegraphics*[width=0.45\textwidth]{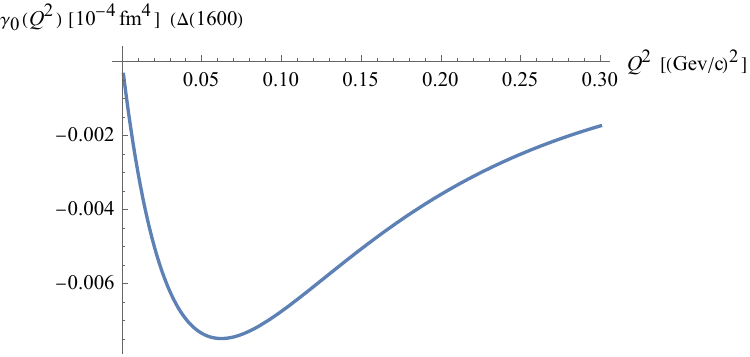}
\end{center}
\caption{\small Contribution to  $\g_{0}(Q^2)$ for the neutron (and proton) from sharp spin 3/2 $\D(1232)$ (left) and $\D(1600)$ nucleon resonance in units of $10^{-4} \text{fm}^4$. }
\label{gnD}
\end{figure}

Negative parity resonances give
\begin{multline}
\g_0(Q^2)_X =  \frac{16 \a_{em} m^4 }{(Q^2 + m_X^2 -m^2)^4}\frac{m_X}{m} \frac{1}{32 M_0^2}(F^{2}_{B_XB}(\vec q\,^{2}) )^2 \ms q^4\times \\\times\big[ |\d_{I_X,I} \d_{n_{\r X}n_\r}-M_0^2\le \r^2\re  \le\tilde  B_X, 1/2| \big(O_1+iO_2\big)|\tilde n,-1/2\re|^2- |M_0^2\le \r^2\re\le \tilde B_X, 3/2| \big(O_1+iO_2\big)|\tilde n,1/2\re|^2\big]\\
=  \frac{ \a_{em} m^4 }{(Q^2 + m_X^2 -m^2)^4}\frac{m_X}{m}  \frac{1}{2 M_0^2}(F^{2}_{B_XB}(\vec q\,^{2}) )^2 Q^4
\bigg[\d_{I_X,I}\bigg(\d_{n_{\r X}n_\r}- \frac43 M_0^2\le \r^2\re\bigg)^2-\d_{I_X,I+1}\frac{32} 9M_0^4 \le\r^2\re^2 \bigg].\\
\end{multline}
In figures \ref{gn}, \ref{gn1} and figure \ref{gnD}, we show the contributions to $\g_{0}(Q^2)$ of the first three low-lying resonances, 
N(1440), N(1535) and $\D(1232)$ and the next excited states N(1710), N(1650) and $\D(1600)$. The $\D(1232)$ state turns out to provide the dominant contribution at low $Q^2$ driving the forward spin polarizability to be negative and increasing with $Q^2$ in qualitative agreement with JLab data \cite{JeffersonLabE97-110:2019fsc}.

\subsubsection{Proton}
The single positive parity resonance contribution to the proton's forward spin polarizability reads
\begin{multline}
\g_0(Q^2)_X =  \frac{16 \a_{em} m^4 }{(Q^2 + m_X^2 -m^2)^4}\frac{m_X}{m} \frac{1}{32 M_0^2}(F^{1}_{B_XB}(\vec q\,^{2}) )^2 \ms q^2\times \\\times\big[ |\d_{I_X,I} \d_{n_{\r X}n_\r}-M_0^2\le \r^2\re  \le \tilde B_X, 1/2| \big(O_1+iO_2\big)|\tilde p,-1/2\re|^2- |M_0^2\le \r^2\re\le \tilde B_X, 3/2| \big(O_1+iO_2\big)|\tilde p,1/2\re|^2\big]\\
=  \frac{ \a_{em} m^4 }{(Q^2 + m_X^2 -m^2)^4}\frac{m_X}{m}  \frac{1}{2 M_0^2}(F^{1}_{B_XB}(\vec q\,^{2}) )^2 Q^2 
\bigg[\d_{I_X,I}\bigg(\d_{n_{\r X}n_\r}+ \frac43 M_0^2\le \r^2\re\bigg)^2-\d_{I_X,I+1}\frac{32} 9M_0^4 \le\r^2\re^2 \bigg].\\
\end{multline}

Negative parity resonances contribute as follows
\begin{multline}
\g_0(Q^2)_X =  \frac{16 \a_{em} m^4 }{(Q^2 + m_X^2 -m^2)^4}\frac{m_X}{m} \frac{1}{32 M_0^2}(F^{2}_{B_XB}(\vec q\,^{2}) )^2 \ms q^4\times \\\times\big[ |\d_{I_X,I}\d_{n_{\r X}n_\r} -M_0^2\le \r^2\re  \le\tilde  B_X, 1/2| \big(O_1+iO_2\big)|\tilde p,-1/2\re|^2- |M_0^2\le \r^2\re\le\tilde  B_X, 3/2| \big(O_1+iO_2\big)|\tilde p,1/2\re|^2\big]\\
=  \frac{ \a_{em} m^4 }{(Q^2 + m_X^2 -m^2)^4}\frac{m_X}{m}  \frac{1}{2 M_0^2}(F^{2}_{B_XB}(\vec q\,^{2}) )^2 Q^4
\bigg[\d_{I_X,I}\bigg(\d_{n_{\r X}n_\r}+ \frac43 M_0^2\le \r^2\re\bigg)^2-\d_{I_X,I+1}\frac{32} 9M_0^4 \le\r^2\re^2 \bigg].\\
\end{multline}

In figure \ref{gp} we show the first low-lying spin 1/2 resonance contributions to proton's $\g_{0}(Q^2)$ at low $Q^2$. The $\Delta(1232)$ contributions, which is isospin-independent is the same as that for the neutron, shown in figure \ref{gnD}. As in the neutron case, the dominant contribution to $\gamma_0(Q^2)$ is given, as expected, by the spin 3/2 $\D(1232)$. The first subdominant contribution appears to be given by the Roper resonance.

\begin{figure}[htb]
\begin{center}
\includegraphics*[width=0.45\textwidth]{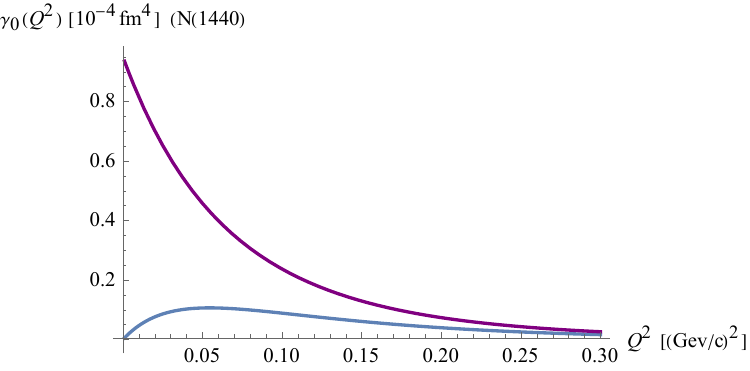}
\includegraphics*[width=0.45\textwidth]{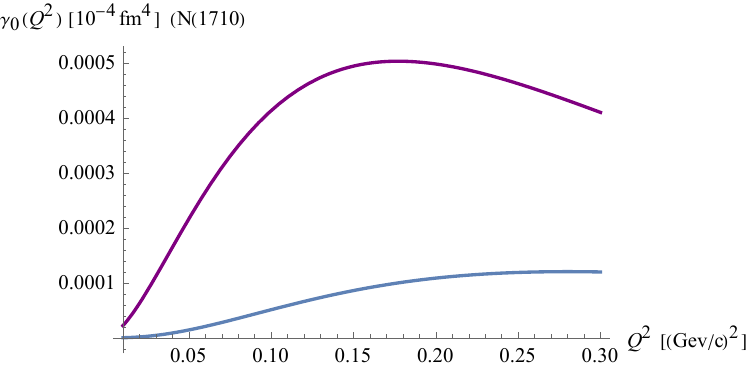}
\end{center}
\caption{\small Contributions to the proton $\g_{0}(Q^2)$ spin polarizability from sharp nucleon resonances N(1440) (left) and N(1710) (right). The solid blue (resp. purple) lines correspond to the non-relativistic (resp. ``relativistic'') result. }
\label{gp}
\end{figure}

\begin{figure}[htb]
\begin{center}
\includegraphics*[width=0.45\textwidth]{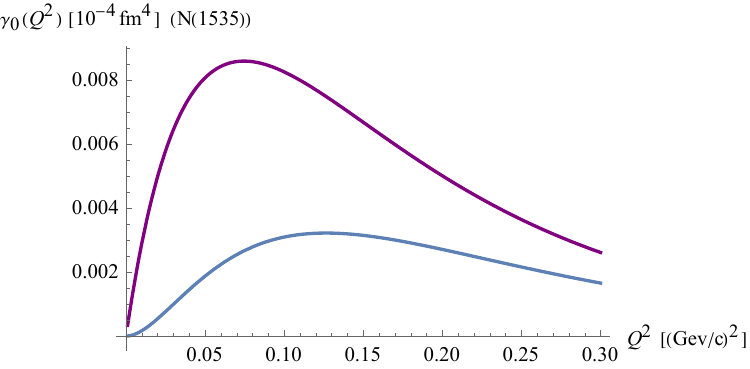}
\includegraphics*[width=0.45\textwidth]{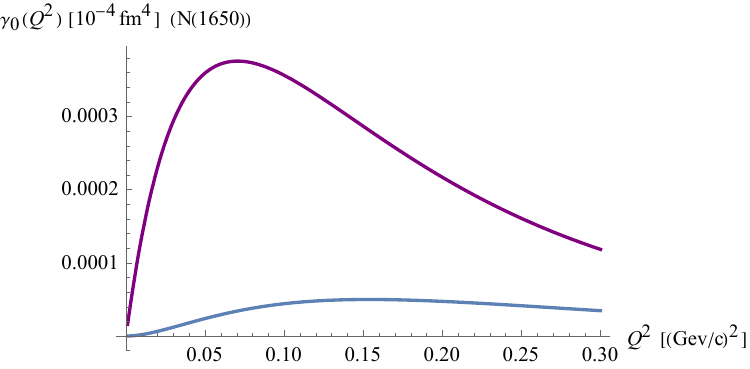}
\end{center}
\caption{\small Contributions to the proton $\g_{0}(Q^2)$ spin polarizability from sharp nucleon resonances N(1535) (left) and N(1650) (right). The solid blue (resp. purple) lines correspond to the non-relativistic (resp. ``relativistic'') result. }
\label{gp}
\end{figure}

\subsection{Total resonance contribution to the spin polarizabilities}
\label{sec:spinpol}
Let us now collect the results obtained in the previous subsections and compute the total contribution of the low-lying resonances to the nucleon spin polarizabilities in the WSS model.

\begin{figure}[htb]
\begin{center}
\includegraphics*[width=0.45\textwidth]{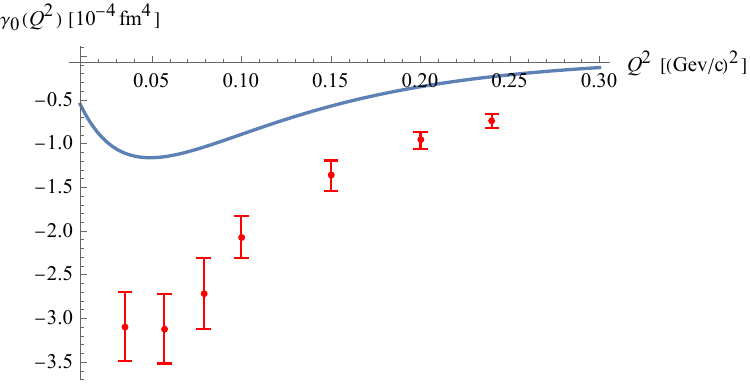}
\includegraphics*[width=0.45\textwidth]{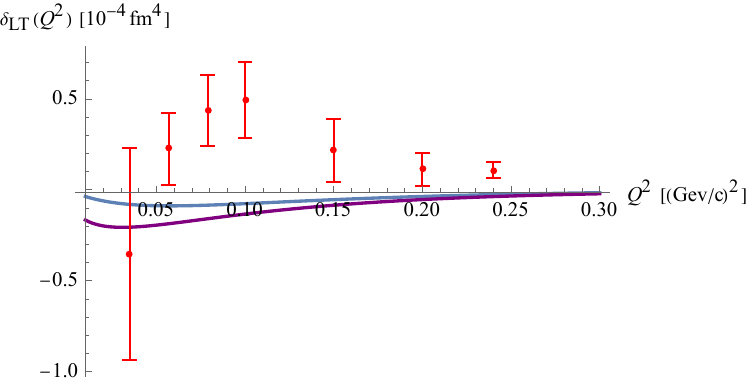}
\end{center}
\caption{\small Contribution to the neutron forward (left) and longitudinal-transverse (right) spin polarizabilities from low-lying sharp nucleon resonances. The solid blue (resp. purple) lines correspond to the non-relativistic (resp. ``relativistic'') result. For $\g_0$, since there is a non-vanishing (and dominant) contribution from the $\D(1232)$, the approximate relativistic expression is not available. The 
results for  $\d_{LT}$ have been multiplied by a factor of 20 to facilitate a qualitative comparison with the data. In red, the experimental data from JLab \cite{E97-110:2021mxm}. }
\label{ntot}
\end{figure}

\begin{figure}[htb]
\begin{center}
\includegraphics*[width=0.45\textwidth]{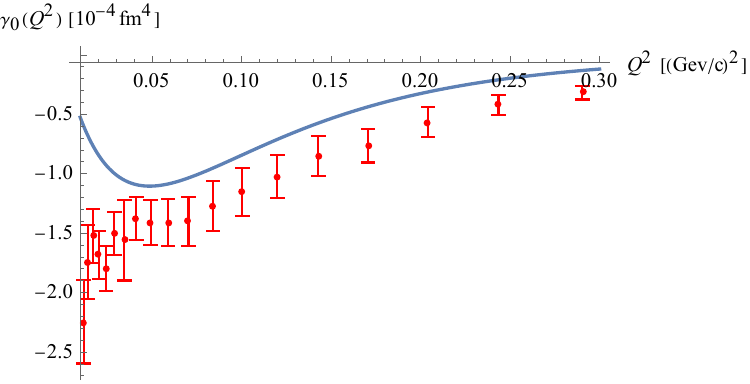}
\includegraphics*[width=0.45\textwidth]{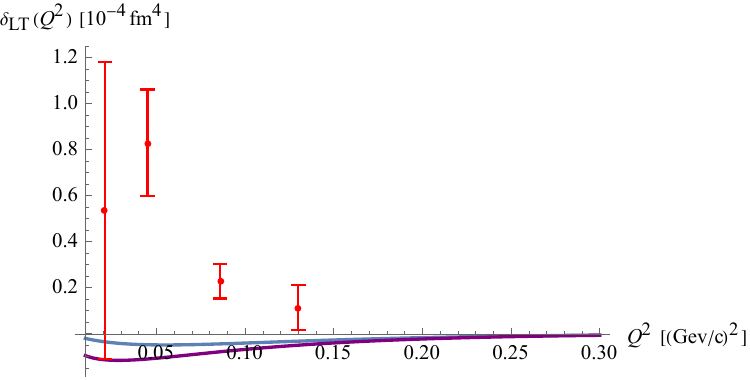}
\end{center}
\caption{\small Contribution to the proton forward (left) and longitudinal-transverse (right) spin polarizabilities from low-lying sharp nucleon resonances. The solid blue and purple lines correspond to results from the non-relativistic and approximate relativistic approaches, respectively. Our findings for the $\d_{LT}$ polarizability have been multiplied by a factor of 10 to be compared with experimental data. Once again, as in the neutron case, for $\g_0$, since there is a non-vanishing (and dominant) contribution from the $\D(1232)$, we do not present the approximate relativistic result. In red, the experimental data from JLab Hall A $g_{2p}$  \cite{JeffersonLabHallAg2p:2022qap} ($\d_{LT}$) and CLAS \cite{CLAS:2021apd} ($\g_0$).  }
\label{ptot}
\end{figure}

A first result of our analysis, valid in the limit of strictly sharp resonances, is presented in Figure \ref{ntot} (for the neutron) and Figure \ref{ptot} (for the proton) where we also report the available experimental data from Jefferson Lab. 

The plots of our results for the forward spin polarizability $\g_0(Q^2)$ (figures \ref{ntot} and \ref{ptot}) do not show evident differences between proton and neutron observables: the reason is that they are dominated by the $\D(1232)$ resonance channel, whose contribution does not depend on the nucleon isospin. As a result, for both nucleons, the total resonance contribution to $\g_0(Q^2)$ is negative and going towards zero as $Q^2$ increases. 
Moreover, our results suggest that, for both nucleons, the resonance contributions to $\d_{LT}(Q^2)$ give rise to a negative function which goes to zero as $Q^2$ increases. 

Comparing our results with realistic data, we must recall that the formers have been actually obtained in the holographic regime of large $N_c$ and large $\lambda$. Actually, $1/\l$ or $1/N_c$ corrections can be relevant in the evaluation of spin-dependent (and other more general) observables when extrapolated to real-world QCD. Moreover, it is important to stress, again, that we have considered only a particular class of contributions to the spin polarizabilities, neglecting for instance the ones from the mesonic cloud, due to pion, vector and axial meson loops. These are also suppressed in the strict holographic regime.

\begin{figure}[htb]
\begin{center}
\includegraphics*[width=0.45\textwidth]{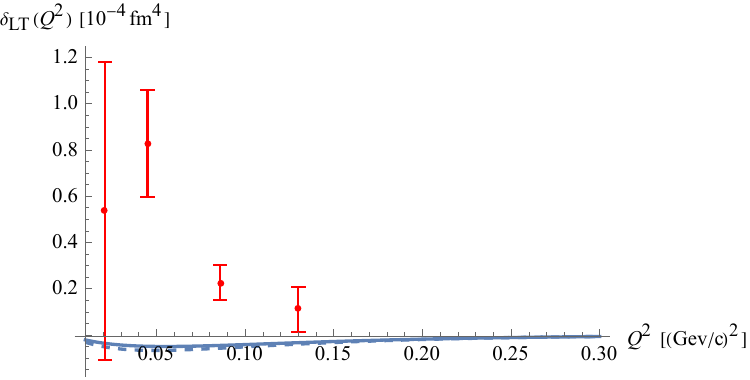}
\includegraphics*[width=0.45\textwidth]{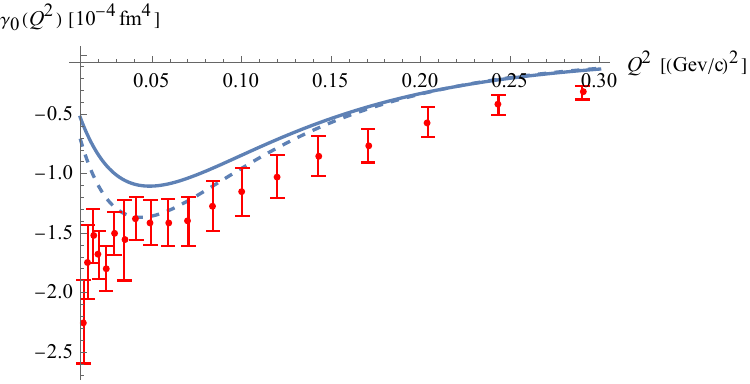}
\end{center}
\caption{\small Contribution to the proton $\d_{LT}(Q^2)$ (left) and $\g_0(Q^2)$ (right) spin polarizabilities from not-sharp low-lying nucleon resonances. The dashed blue (resp. solid blue) lines correspond to the non-relativistic not-sharp (resp. sharp) result. Our results for $\d_{LT}(Q^2)$ are shown multiplied by a factor of 10 to be better compared with the data. In red, the experimental data from JLab Hall A $g2p$  \cite{JeffersonLabHallAg2p:2022qap} ($\d_{LT}$) and CLAS \cite{CLAS:2021apd} ($\g_0$). }
\label{deltaLTnotstable}
\end{figure}

\begin{figure}[htb]
\begin{center}
\includegraphics*[width=0.45\textwidth]{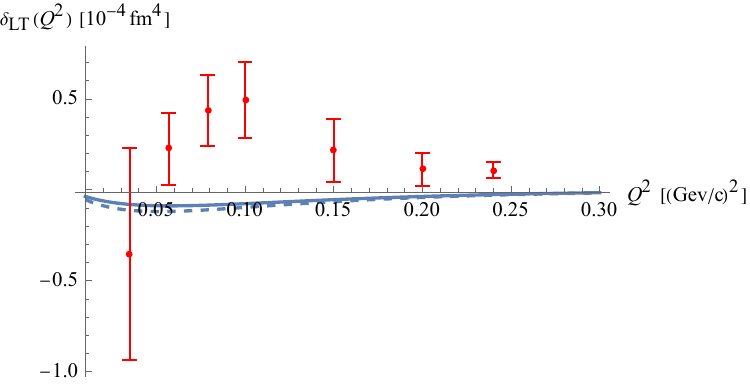}
\includegraphics*[width=0.45\textwidth]{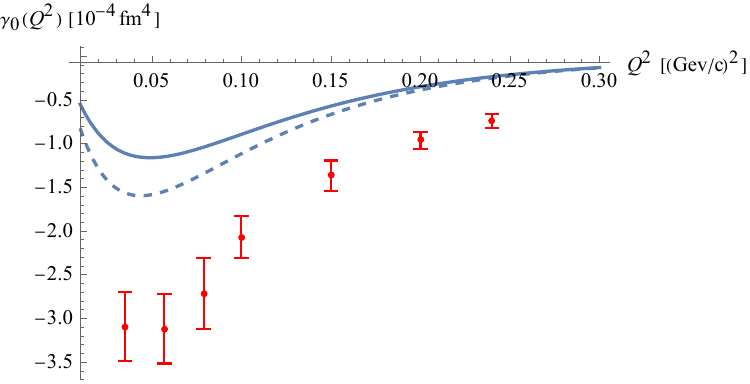}
\end{center}
\caption{\small Contribution to the neutron $\d_{LT}(Q^2)$ and $\g_0(Q^2)$ spin polarizabilities from not-sharp low-lying nucleon resonances. The dashed blue (resp. solid blue ) lines correspond to the non-relativistic not-sharp (resp. sharp) result. In order to have a better comparison, our results for the $\d_{LT}(Q^2)$ polarizability are presented here multiplied by a factor of 20. In red, the experimental data from JLab \cite{E97-110:2021mxm}.}
\label{deltaLTnotstablen}
\end{figure}

In addition, till now, we have treated the nucleon resonances as stable particles: this is unavoidable in the large $N_c$ limit in the WSS model but it is interesting to consider, at a more phenomenological level, the realistic possibility of having non-sharp resonances. In this case, the equations (\ref{carlson}) for the structure functions can be modified as suggested in \cite{Carlson:1998gf}, i.e. replacing the Dirac delta functions for sharp resonances with the following approximation 
\be
\d((p+q)^2 + m_X^2) \to \frac{1}{4\pi m_X} \frac{\G_X}{\big(\sqrt{|(p+q)^2| }- m_X\big)^2 + \G_X^2/4}\,,
\ee
where $\G_X$ is the width of the resonance (taken as an input from the experimental data \cite{ParticleDataGroup:2020ssz}). Then, using the modified expressions for $g_{1}(x,Q^2)$ and $g_{2}(x,Q^2)$ and the definitions (\ref{expgamma0}), (\ref{expdeltaLT}) we can obtain an estimate for the not-stable resonances contribution to the forward and longitudinal spin polarizabilities
\be
\label{expgamma01}
\gamma_0(Q^2) = \frac{4\alpha_{em} m^4}{\pi Q^6}\sum_{m_X} \frac{\G_X}{m_X}  \big[| G^{+}_{B_XB}(Q^2)|^2-| G^{-}_{B_XB}(Q^2)|^2\big] \mc F_1(Q^2, m_X, \G_X),
\ee
\be
\label{expdeltaLT1}
\delta_{LT}(Q^2) = \frac{2\sqrt 2 \alpha_{em} m^3 }{\pi Q^5}\sum_{m_X} (-1)^{S_X-1/2}\eta_X\frac{\G_X}{m_X}   G^{+\,*}_{B_XB}(Q^2) G^{0}_{B_XB}(Q^2) \mc F_2(Q^2, m_X, \G_X),
\ee
with $\mc F_{1,2}(Q^2, m_X, \G_X)$ given by
\ba
&&\mc F_1(Q^2, m_X, \G_X) =  \int_{0}^{x_0} dx\, \frac{x^2}{\big(\sqrt{|(p+q)^2| }- m_X\big)^2 + \G_X^2/4}, \nb\\
&&\mc F_2(Q^2, m_X, \G_X)= \int_{0}^{x_0} dx\, \frac{x}{\big(\sqrt{|(p+q)^2| }- m_X\big)^2 + \G_X^2/4}\,.
\ea
Let us adopt a simple approximate approach and look at what happens by evaluating (\ref{expgamma01}) and (\ref{expdeltaLT1}) still using the expressions in (\ref{G+}), (\ref{G0}) and \ref{G-}) for the helicity amplitudes. Following this prescription, in figures \ref{deltaLTnotstable} and \ref{deltaLTnotstablen} we show the low-$Q^2$ behavior of the spin polarizabilities for proton and neutron respectively.

We can thus notice that in the non-sharp case the absolute values of the polarizabilities are only very slightly increased with respect to the sharp case. 
\subsubsection{Resonance contributions: data interpolation}

\begin{figure}[htb]
\begin{center}
\includegraphics*[width=0.45\textwidth]{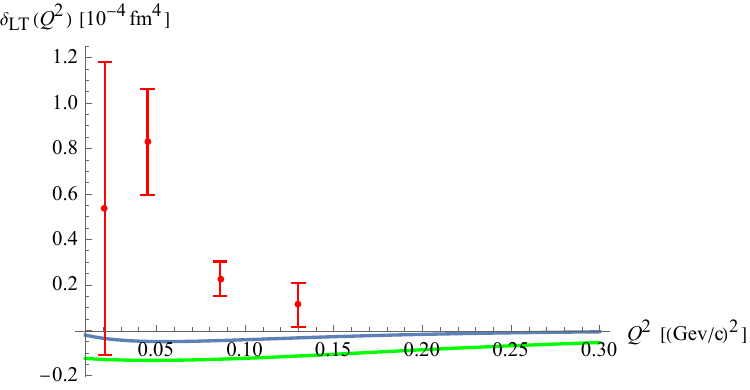}
\includegraphics*[width=0.45\textwidth]{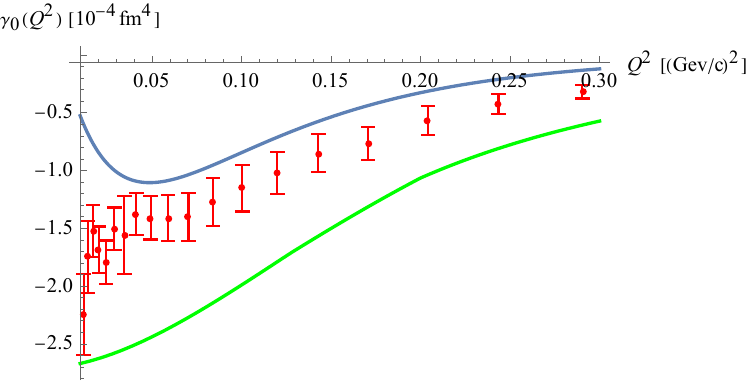}
\end{center}
\caption{\small Contribution to the proton $\d_{LT}(Q^2)$ and $\g_0(Q^2)$ spin polarizabilities form resonances only. The solid blue lines correspond to our results in the case of non-relativistic sharp resonances. The green lines are derived from data interpolation of helicity amplitudes for $\g\, p \to N^*$ transition, as explained in the main text. Our results for the $\d_{LT}(Q^2)$ polarizability are presented here multiplied by a factor of 10. In red, the experimental data from Jefferson Lab Collaboration E97-110 \cite{E97-110:2021mxm} and Hall A $g_{2p}$\cite{JeffersonLabHallAg2p:2022qap}.}
\label{dataint}
\end{figure}

With the aim of investigating the reliability of the WSS model, it is useful to compare our results not just with the total measured polarizabilities, but, better, with the contributions to the latter due to the resonances alone. In order to carry out this check, let us focus on the helicity amplitudes for the proton-resonance transitions where data at $Q^2>0$ are available. Following what has been done for the contributions of the resonances to the polarized structure functions in \cite{HillerBlin:2022ltm} and making use of the JLab results reported therein, we interpolate the experimental data for $A_{1/2}(Q^2)$,  $S_{1/2}(Q^2)$ and eventually $A_{3/2}(Q^2)$ for the low-lying nucleon resonances. Subsequently, using the expressions in (\ref{defhelamp}), (\ref{rescontd}) and (\ref{rescontg}), we estimate the expected contributions of the resonances to the proton spin polarizabilities. The results are shown in green in figure \ref{dataint} and are in qualitative agreement with the holographic ones.

\section{Conclusions}
\label{sec:conc}
The recent experimental results from the Jefferson Lab Collaboration E97-110 \cite{E97-110:2021mxm} and Hall A $g_{2p}$\cite{JeffersonLabHallAg2p:2022qap} on the longitudinal-transverse ($\d_{LT}(Q^2)$) and forward ($\g_0(Q^2)$) generalized nucleon spin polarizabilities pose an extremely relevant challenge to the current theoretical treatment of the nucleon spin structure. The measurements at low energies of the latter observables, describing the precession of the nucleon composite spin structure in response to external electromagnetic fields, show, especially in the case of the neutron $\delta_{LT}(Q^2)$, qualitative and quantitative differences with the available theoretical predictions. 
These discrepancies leave behind an unresolved and incomplete description of the nucleon spin structure. 

In light of these issues, this work aimed, as a first step, to study the resonance contributions to the nucleon spin structure functions at low-energy by means of a non-perturbative tool based on the holographic correspondence, focusing on the Witten-Sakai-Sugimoto (WSS) model \cite{Witten:1998zw,Sakai:2004cn}. The latter is a large $N_c$, strongly coupled QCD-like theory with many low-energy features common to QCD, such as confinement and chiral symmetry breaking. Extrapolating the model parameters to realistic QCD data, many WSS observables have been computed in the literature with ${\cal O}(10\%-20\%)$ accuracy. Most importantly, the model offers the advantage to get insights on features, such as the baryonic resonance contributions to the nucleon spin structure functions, which are otherwise difficult to access using more standard theoretical approaches. 

The strategy followed in this work has been the following.
The polarized electron-nucleon scattering processes, considered in \cite{E97-110:2021mxm, JeffersonLabHallAg2p:2022qap}  to extract the experimental data on $\d_{LT}(Q^2)$ and $\g_0(Q^2)$, are described by a cross-section which depends on the so-called hadronic tensor $W^{\m\n}$, containing the whole information of the hadronic part of the process. The latter can be expressed as a sum, over all possible final hadronic states, of products of matrix elements of the electromagnetic current evaluated between the initial nucleonic state and the final hadronic state $X$. These are in turn related to the nucleon-resonance helicity amplitudes. 
Moreover, $W^{\m\n}$ can be decomposed in terms of the non-polarized $F_1(x,Q^2)$ and $F_2(x,Q^2)$ and polarized $g_1(x,Q^2)$ and $g_2(x,Q^2)$  structure functions, which encode the internal spin-independent and spin-dependent structure of the target nucleon. Computing the helicity amplitudes it is possible to reconstruct the resonance contributions to the structure functions, using the general relations found in \cite{Carlson:1998gf}. From the moments of two specific combinations of $g_1(x,Q^2)$ and $g_2(x,Q^2)$, it is then possible to derive the expressions for $\d_{LT}(Q^2)$ and $\g_0(Q^2)$.

In particular, following \cite{Hata:2007mb}, we have analyzed the baryon states which arise as instanton configurations of the low energy 5d effective action of the WSS model. Then, using the prescription in \cite{Hashimoto:2008zw}, we have introduced the electromagnetic current in the model and we have computed its matrix elements between the initial nucleon state and a final baryonic state $X$, focusing, for the latter, on low-lying resonances with spin $3/2$ and spin $1/2$ with positive and negative parity. 

Collecting our results we have extracted the helicity amplitudes and then the resonance contributions to the nucleon spin polarizabilities. 
Our analysis, with WSS parameters extrapolated to QCD data, shows that the spin $3/2$ $\Delta(1232)$ resonance gives the dominant contribution to the forward spin polarizabilities $\gamma_0(Q^2)$. The contribution is isospin-independent and drives the polarizabilities to negative values at low $Q^2$ in qualitative agreement with experimental results. The contribution of spin $1/2$ resonances to nucleons' $\gamma_0(Q^2)$ is always positive.

The contribution of the $\Delta(1232)$ resonance to the $\delta_{LT}$ polarizabilities is instead negligible (it is actually zero to leading order in the holographic limit in the WSS model): this is also in agreement with JLab observations. The low-lying spin $1/2$ resonances $N(1440)$, $N(1710)$ (with positive parity) and $N(1535)$, $N(1650)$ (with negative parity) provide different contributions to the longitudinal-transverse polarizabilities. Actually (with the exception the $N(1440)$ whose contribution is zero for both nucleons in the WSS model), spin $1/2$ positive parity resonances give positive (and asymptotically decreasing) contributions to proton's $\delta_{LT}(Q^2)$ while they do not contribute to neutron's one. Negative parity spin $1/2$ resonances, instead, give rise to negative contributions, going towards zero as $Q^2$ increases. For both nucleons, the sum of the various contributions produces a negative $\delta_{LT}(Q^2)$ function, asymptotically decreasing with $Q^2$ in absolute value. These features, at least in the proton case where enough data are available, qualitatively agree with those extrapolated from experimental results on the helicity amplitudes at low $Q^2$. 

Our analysis suggests that resonances alone are not enough to catch the low $Q^2$ behavior of the polarizabilities, especially for what concerns the longitudinal-transverse ones. This suggests that other (meson cloud) contributions, which are not accounted for by our analysis, might play a relevant role in the game.

This work, which provides the first study of polarized lepton-nucleon scattering at low energies through the WSS holographic model, opens the way to several interesting and challenging developments. Future possible directions include looking for a relativistic shift for the spin $3/2$ states in the WSS model and adding quark masses and isospin breaking effects to the analysis. Moreover, following a complementary approach with respect to that adopted in the present work, it will be important to compute the hadronic tensor $W^{\mu\nu}$ having in mind its expression as written in formula (\ref{formula2}), i.e. via  a direct holographic computation of the electromagnetic current two-point functions. Through the latter approach, it could be possible to have a more complete picture of the scattering process, accounting for different possible hadronic contributions. Work is in progress in this direction \cite{inprogress}. Finally it might be important to include in the analysis possible anomaly contributions to the nucleon spin polarizabilities, along the lines of \cite{Kochelev:2011bh} and to holographically compute the various sum rules (see e.g. \cite{Deur:2018roz}) which are expected to be satisfied by (combinations of) $g_1, g_2$ and their moments. 

\vskip 15pt \centerline{\bf Acknowledgments} \vskip 10pt 

\noindent 
We are deeply indebted to A.L. Cotrone for suggestions, comments and discussions and for bringing to our attention the experimental results in \cite{E97-110:2021mxm}. We thank A.L. Cotrone and  J. Leutgeb for their collaboration in the early stages of this work and F. Becattini for comments. 
\appendix
\section{Baryon Wave-functions}
\label{appa}
Here we summarize the wave-functions for the quantized low-lying baryons, mainly referring to Appendix A of \cite{Hashimoto:2008zw}.
Recall that in the WSS model, baryon eigenstates are labelled by the quantum numbers $B\equiv  (l,I_3,n_\r,n_z)$ and spin component $s$ (with $S= I = l/2$). Moreover, they satisfy the property $\psi(\bf{a}) = -\psi(-{\bf{a}})$, so the wave-functions will be polynomials of odd degrees in the ${\bf{a}}$ coordinates. For instance, states with $I=S=1/2$ correspond to wave-functions linear in ${\bf{a}}$ and states with $I=S=3/2$ to cubic functions. Here it is useful to bring to mind the expressions of the spin and isospin operators in the ${\bf{a}}$ space:
\be
\label{IS}
I_a = \frac{i}{2}\bigg(a_4\frac{\p}{\p a_a}- a_a\frac{\p}{\p a_4} - \e_{abc}a_b\frac{\p}{\p a_c}\bigg)\,, \hspace{0.5cm} S_a = \frac{i}{2}\bigg(-a_4\frac{\p}{\p a_a}+ a_a\frac{\p}{\p a_4} - \e_{abc}a_b\frac{\p}{\p a_c}\bigg)\,.
\ee
Nucleon states are specified by 
\be
B \equiv  (1,\pm1/2, 0,0), \hspace{1cm} s= \pm1/2\,,
\ee
and the corresponding wave-functions are written as
\ba
&&\vert p, \vec p , s= +1/2\rangle =  \frac{1}{(2\pi)^{3/2}}e^{i\vec {p} \cdot \vec {X}}R_{0(1)}(\r)\psi_0(Z)\bigg(\frac{1}{\pi} (a_1+ia_2)\bigg),\nb\\
&&\vert p, \vec p , s= -1/2\rangle = \frac{1}{(2\pi)^{3/2}}e^{i\vec {p} \cdot \vec {X}}R_{0(1)}(\r)\psi_0(Z)\bigg(-\frac{i}{\pi} (a_4-ia_3)\bigg),\nb\\
&&\vert n, \vec p , s= +1/2\rangle =   \frac{1}{(2\pi)^{3/2}}e^{i\vec {p} \cdot \vec {X}}R_{0(1)}(\r)\psi_0(Z) \bigg(\frac{i}{\pi} (a_4+ia_3)\bigg),\nb\\
&&\vert n, \vec p , s= -1/2\rangle =  \frac{1}{(2\pi)^{3/2}}e^{i\vec {p} \cdot \vec {X}}R_{0(1)}(\r)\psi_0(Z) \bigg(-\frac{1}{\pi} (a_1-ia_2)\bigg),
\ea
where $R_{0(1)}(\r)$ and $\psi_0(Z)$ read
\be
\label{wfnucleon}
R_{0(1)}(\r) = e^{-\frac{M_0}{\sqrt{6}}\r^2}\r^{-1 + 2\sqrt{ 1+N_c^2/5}}\,, \hspace{1cm} \psi_{0}(Z) = \bigg( \frac{2M_0}{\sqrt{6}\pi^2}\bigg)^{1/4} e^{-\frac{M_0}{\sqrt{6}}Z^2}\,.
\ee
Then, let us consider the first excited states. The baryon state with quantum numbers 
\be
\label{roperqn}
B= (1,\pm1/2,1,0), \hspace{1cm} s= \pm 1/2\,,
\ee
corresponds to the Roper excitation $N(1440)$, $I(S^P) = 1/2(1/2^+)$, and its wave-function can be obtained by replacing in the nucleon wave-function $R_{0(1)} \to R_{1(1)}$, where
\ba
\label{roper}
R_{1(1)} &=& e^{-\frac{M_0}{\sqrt{6}}\r^2}\r^{-1 + 2\sqrt{ 1+N_c^2/5}} \, _1F_1(-1, 1 + 2\sqrt{1 +N_c^2/5}, 2/\sqrt{6} M_0\r^2)\nb\\
&=& \bigg(\frac{2M_0}{\sqrt{6}}\r^2 -1-2\sqrt{ 1+N_c^2/5}\bigg)e^{-\frac{M_0}{\sqrt{6}}\r^2}\r^{-1 + 2\sqrt{ 1+N_c^2/5}}\,.
\ea

Another interesting excitation to analyze is $N(1535)$, $I(S^P) = 1/2(1/2^-)$: this is identified with the first excited baryon state in the $Z$ part of the spectrum, i.e. 
\be
B= (1,\pm1/2,0,1)\,, \hspace{1cm} s= \pm 1/2\,.
\ee
The corresponding wave-function is the same as that of the neutron and proton, except for the $Z$ eigenfunction which is now given by
\ba
\label{wfzn1353}
\psi_1 (Z) &=& \frac{1}{2} \bigg( \frac{2M_0}{\sqrt{6}\pi^2}\bigg)^{1/4} H_{1} \big(\sqrt{2M_0}6^{-1/4} Z\big)e^{-\frac{M_0}{\sqrt{6}}Z^2}\nb\\
&=&  \bigg( \frac{2M_0}{\sqrt{6}\pi^2}\bigg)^{1/4} 2 \sqrt{M_0}6^{-1/4} Z e^{-\frac{M_0}{\sqrt{6}}Z^2}.
\ea
This is clearly odd under the transformation $Z \to -Z$. This property prescribes the oddness of the state under the parity transformation in the WSS model \cite{Sakai:2004cn}, in agreement with the identification of the above state with the negative parity nucleon excitation $N(1535)$.

The first excited states that we encounter in the $SU(2)$ sector are the ones with quantum numbers $l= 3$: these states realize baryons with $I=S=3/2$. The lightest among these excited states ($n_\r = n_z=0$) can be identified with the $\D(1232)$ baryons:  $\D^{++}, \, \D^{+}, \, \D^{0},$ and $\D^{-}$, depending on the $I_3$ value. In particular, the $SU(2)$ part of the wave-function of the higher isospin state $\D^{++}$, in the case $s= 3/2$, is given by \cite{Adkins:1983ya}
\be
\label{delta++}
\vert \D^{++}, s= 3/2\rangle \propto (a_1 + ia_2)^3\,.
\ee
Then, the other wave-functions with $s=3/2$ can be obtained by acting on (\ref{delta++}) with the isospin lowering operator:
\ba
&&\vert \D^{+}, s= 3/2\rangle \propto (a_1 + ia_2)^2 (a_4+ia_3)\,,\nb\\
&&\vert \D^{0}, s= 3/2\rangle \propto (a_1 + ia_2) (a_4+ia_3)^2\,,\nb\\
&&\vert \D^{+}, s= 3/2\rangle \propto (a_4+ia_3)^3\,.
\ea
Among the above wavefunctions, it is convenient to list the non-relativistic spin-isospin parts of the $\D^0$ and $\D^+$ states (up to an overall phase factor):
\ba
&&|\D^0, s = 3/2\re = \frac{\sqrt 6}{\pi} (a_1 + i a_2)(a_4 + i a_3)^2\,,\nb\\
&&|\D^0, s = 1/2\re =  i\frac{\sqrt 2}{\pi} (1-3a_1^2 - 3a_2^2)(a_4 + i a_3)\,,\nb\\
&&|\D^0, s = -1/2\re =-\frac{\sqrt 2}{\pi}(1-3a_3^2 - 3a_4^2) (a_1 - i a_2)\,,\nb\\
&&|\D^0, s = -3/2\re = i\frac{\sqrt 6}{\pi} (a_1 - i a_2)^2(a_4 - i a_3)\,.
\ea
Those states will be useful in evaluating the amplitude of the $\g\, n \to \D^0$  transition. Despite the $\g\, p \to \D^+$ transition would give the same results, for amplitudes and electromagnetic current matrix elements, as the $\g\, n \to \D^0$  one,  for completeness we list also the non-relativistic spin-isospin parts of the $\D^+$ states:
\ba
&&|\D^+, s = 3/2\re = \frac{\sqrt 6}{\pi} (a_1 + i a_2)^2(a_4 + i a_3)\,,\nb\\
&&|\D^+, s = 1/2\re =  i\frac{\sqrt 2}{\pi} (1-3a_3^2 - 3a_4^2)(a_1 + i a_2)\,,\nb\\
&&|\D^+, s = -1/2\re =-\frac{\sqrt 2}{\pi}(1-3a_1^2 - 3a_2^2) (a_4 - i a_3)\,,\nb\\
&&|\D^+, s = -3/2\re = -i\frac{\sqrt 6}{\pi} (a_1-i  a_2)(a_4 - i a_3)^2\,.
\ea
The wavefunctions corresponding to the $\D(1232)$ baryons are completed by considering the ground state eigenfunctions in the $\r$ and $Z$ sectors, i.e. $R_{0(3)} (\r) = R_{0(1)}(\r)$ and $\psi_0(Z)$.

Finally, let us collect here some useful relations related to the operator (\ref{Oa}):
\ba
\label{o1o2}
 \le N_X, 1/2| \big(O_2 -iO_1\big)|N,-1/2\re = \t^3_{I^3_X,I^3} \frac{4i}3\,,\nb\\
 \le \D, 1/2| \big(O_2 -iO_1\big)|N,-1/2\re= -2\d_{I^3_X,I^3} \frac{\sqrt 2}3\,,\nb\\
  \le \D, 3/2| \big(O_2 -iO_1\big)|N,1/2\re= -2\d_{I^3_X,I^3} \sqrt{\frac23}\,.
\ea


\begin{thebibliography}{99}
 \bibitem{Jaffe:1996zw}
R.~L.~Jaffe,
``Spin, twist and hadron structure in deep inelastic processes,''
[arXiv:hep-ph/9602236 [hep-ph]].
\bibitem{Deur:2018roz}
A.~Deur, S.~J.~Brodsky and G.~F.~De T\'eramond,
``The Spin Structure of the Nucleon,''
[arXiv:1807.05250 [hep-ph]].
\bibitem{Manohar:1992tz}
A.~V.~Manohar,
``An Introduction to spin dependent deep inelastic scattering,''
[arXiv:hep-ph/9204208 [hep-ph]].
\bibitem{Drechsel:2002ar}
D.~Drechsel, B.~Pasquini and M.~Vanderhaeghen,
``Dispersion relations in real and virtual Compton scattering,''
Phys. Rept. \textbf{378}, 99-205 (2003)
[arXiv:hep-ph/0212124 [hep-ph]].

\bibitem{JeffersonLabE97-110:2019fsc}
V.~Sulkosky \textit{et al.} [Jefferson Lab E97-110],
``Measurement of the 3He spin-structure functions and of neutron (3He) spin-dependent sum rules at $0.035\leq Q^2\leq 0.24$ $GeV^2$,"
Phys. Lett. B \textbf{805}, 135428 (2020)
[arXiv:1908.05709 [nucl-ex]].

\bibitem{E97-110:2021mxm}
V.~Sulkosky \textit{et al.} [E97-110],
``Puzzle with the precession of the neutron spin,''
Nature Phys. \textbf{17}, no.6, 687-692 (2021)
[arXiv:2103.03333 [nucl-ex]].
\bibitem{Deur:2022sot}
A.~Deur,
``Results on spin sum rules and polarizabilities at low $Q^2$,''
[arXiv:2202.10511 [nucl-ex]].
\bibitem{Bernard:2002pw}
V.~Bernard, T.~R.~Hemmert and U.~G.~Meissner,
``Spin structure of the nucleon at low-energies,''
Phys. Rev. D \textbf{67}, 076008 (2003)
[arXiv:hep-ph/0212033 [hep-ph]].
\bibitem{Kao:2002cp}
C.~W.~Kao, T.~Spitzenberg and M.~Vanderhaeghen,
``Burkhardt-Cottingham sum rule and forward spin polarizabilities in heavy baryon chiral perturbation theory,''
Phys. Rev. D \textbf{67}, 016001 (2003)
[arXiv:hep-ph/0209241 [hep-ph]].

\bibitem{Bernard:2012hb}
V.~Bernard, E.~Epelbaum, H.~Krebs and U.~G.~Meissner,
``New insights into the spin structure of the nucleon,''
Phys. Rev. D \textbf{87}, no.5, 054032 (2013)
[arXiv:1209.2523 [hep-ph]].

\bibitem{Lensky:2014dda}
V.~Lensky, J.~M.~Alarc\'on and V.~Pascalutsa,
``Moments of nucleon structure functions at next-to-leading order in baryon chiral perturbation theory,''
Phys. Rev. C \textbf{90}, no.5, 055202 (2014)
[arXiv:1407.2574 [hep-ph]].

\bibitem{Alarcon:2020icz}
J.~M.~Alarc\'on, F.~Hagelstein, V.~Lensky and V.~Pascalutsa,
``Forward doubly-virtual Compton scattering off the nucleon in chiral perturbation theory: II. Spin polarizabilities and moments of polarized structure functions,''
Phys. Rev. D \textbf{102}, no.11, 114026 (2020)
[arXiv:2006.08626 [hep-ph]].
\bibitem{Drechsel:2000ct}
D.~Drechsel, S.~S.~Kamalov and L.~Tiator,
``The GDH sum rule and related integrals,''
Phys. Rev. D \textbf{63}, 114010 (2001)
[arXiv:hep-ph/0008306 [hep-ph]].

\bibitem{JeffersonLabHallAg2p:2022qap}
D.~Ruth \textit{et al.} [Jefferson Lab Hall A g2p],
``Proton spin structure and generalized polarizabilities in the strong quantum chromodynamics regime,''
Nature Phys. \textbf{18}, no.12, 1441-1446 (2022)
[arXiv:2204.10224 [nucl-ex]].

\bibitem{CLAS:2021apd}
X.~Zheng \textit{et al.} [CLAS],
``Measurement of the proton spin structure at long distances,''
Nature Phys. \textbf{17} (2021) no.6, 736-741
[arXiv:2102.02658 [nucl-ex]].

\bibitem{Witten:1998zw}
E.~Witten,
``Anti-de Sitter space, thermal phase transition, and confinement in gauge theories,''
Adv. Theor. Math. Phys. \textbf{2}, 505-532 (1998)
[arXiv:hep-th/9803131 [hep-th]].
\bibitem{Sakai:2004cn}
T.~Sakai and S.~Sugimoto,
``Low energy hadron physics in holographic QCD,''
Prog. Theor. Phys. \textbf{113}, 843-882 (2005)
[arXiv:hep-th/0412141 [hep-th]].
\bibitem{Adkins:1983ya}
G.~S.~Adkins, C.~R.~Nappi and E.~Witten,
``Static Properties of Nucleons in the Skyrme Model,''
Nucl. Phys. B \textbf{228} (1983), 552
\bibitem{Hata:2007mb}
H.~Hata, T.~Sakai, S.~Sugimoto and S.~Yamato,
``Baryons from instantons in holographic QCD,''
Prog. Theor. Phys. \textbf{117}, 1157 (2007)
[arXiv:hep-th/0701280 [hep-th]].
\bibitem{Hashimoto:2008zw}
K.~Hashimoto, T.~Sakai and S.~Sugimoto,
``Holographic Baryons: Static Properties and Form Factors from Gauge/String Duality,''
Prog. Theor. Phys. \textbf{120}, 1093-1137 (2008)
[arXiv:0806.3122 [hep-th]].
\bibitem{Bayona:2011xj}
C.~A.~B.~Bayona, H.~Boschi-Filho, N.~R.~F.~Braga, M.~Ihl and M.~A.~C.~Torres,
``Generalized baryon form factors and proton structure functions in the Sakai-Sugimoto model,''
Nucl. Phys. B \textbf{866}, 124-156 (2013)
[arXiv:1112.1439 [hep-ph]].
\bibitem{Ballon-Bayona:2012txi}
A.~Ballon-Bayona, H.~Boschi-Filho, N.~R.~F.~Braga, M.~Ihl and M.~A.~C.~Torres,
``Production of negative parity baryons in the holographic Sakai-Sugimoto model,''
Phys. Rev. D \textbf{86}, 126002 (2012)
[arXiv:1209.6020 [hep-ph]].
\bibitem{Grigoryan:2009pp}
H.~R.~Grigoryan, T.~S.~H.~Lee and H.~U.~Yee,
``Electromagnetic Nucleon-to-Delta Transition in Holographic QCD,''
Phys. Rev. D \textbf{80} (2009), 055006
[arXiv:0904.3710 [hep-ph]].

\bibitem{Carlson:2003je}
C.~E.~Carlson and C.~R.~Ji,
``Angular conditions, relations between Breit and light front frames, and subleading power corrections,''
Phys. Rev. D \textbf{67}, 116002 (2003)
[arXiv:hep-ph/0301213 [hep-ph]].


\bibitem{Carlson:1998gf}
C.~E.~Carlson and N.~C.~Mukhopadhyay,
``Bloom-Gilman duality in the resonance spin structure functions,''
Phys. Rev. D \textbf{58} (1998), 094029
[arXiv:hep-ph/9801205 [hep-ph]].
\bibitem{Ramalho:2019ocp}
G.~Ramalho,
``Low-$Q^2$ empirical parametrizations of the $N^\ast$ helicity amplitudes,''
Phys. Rev. D \textbf{100} (2019) no.11, 114014
[arXiv:1909.00013 [hep-ph]].
\bibitem{Aznauryan:2008us}
I.~G.~Aznauryan, V.~D.~Burkert and T.~S.~H.~Lee,
``On the definitions of the gamma* N ---\ensuremath{>} N* helicity amplitudes,''
[arXiv:0810.0997 [nucl-th]].

\bibitem{Ramalho:2023hqd}
G.~Ramalho and M.~T.~Pe\~na,
``Electromagnetic Transition Form Factors of Baryon Resonances,''
[arXiv:2306.13900 [hep-ph]].

\bibitem{Aharony:2008an}
O.~Aharony and D.~Kutasov,
``Holographic Duals of Long Open Strings,''
Phys. Rev. D \textbf{78}, 026005 (2008)
[arXiv:0803.3547 [hep-th]].
\bibitem{Hashimoto:2008sr}
K.~Hashimoto, T.~Hirayama, F.~L.~Lin and H.~U.~Yee,
``Quark Mass Deformation of Holographic Massless QCD,''
JHEP \textbf{07}, 089 (2008)
[arXiv:0803.4192 [hep-th]].
\bibitem{sheet}
F.~Bigazzi, A.~L.~Cotrone and A.~Olzi,
``Hall Droplet Sheets in Holographic QCD,''
JHEP \textbf{02}, 194 (2023)
[arXiv:2211.05147 [hep-th]].

\bibitem{komar} 
Z.~Komargodski,
``Baryons as Quantum Hall Droplets,''
[arXiv:1812.09253 [hep-th]].

\bibitem{Imaanpur:2022lvp}
A.~Imaanpur,
``Correction to baryon spectrum in holographic QCD,''
Phys. Lett. B \textbf{832}, 137233 (2022)
[arXiv:2206.13878 [hep-th]].

\bibitem{Bigazzi:2018cpg}
F.~Bigazzi and P.~Niro,
``Neutron-proton mass difference from gauge/gravity duality,''
Phys. Rev. D \textbf{98} (2018) no.4, 046004
[arXiv:1803.05202 [hep-th]].

\bibitem{Fujii:2022yqh}
D.~Fujii, A.~Iwanaka and A.~Hosaka,
``Electromagnetic transition amplitude for Roper resonance from holographic QCD,''
Phys. Rev. D \textbf{106} (2022) no.1, 014010
[arXiv:2203.13988 [hep-ph]].

\bibitem{Blin:2021twt}
A.~N.~H.~Blin, W.~Melnitchouk, V.~I.~Mokeev, V.~D.~Burkert, V.~V.~Chesnokov, A.~Pilloni and A.~P.~Szczepaniak,
``Resonant contributions to inclusive nucleon structure functions from exclusive meson electroproduction data,''
Phys. Rev. C \textbf{104} (2021) no.2, 025201
[arXiv:2105.05834 [hep-ph]].

\bibitem{Boschi-Filho:2011jen}
H.~Boschi-Filho, N.~R.~F.~Braga, M.~Ihl and M.~A.~C.~Torres,
``Relativistic baryons in the Skyrme model revisited,''
Phys. Rev. D \textbf{85} (2012), 085013
[arXiv:1111.2287 [hep-th]].

\bibitem{CLAS:2009tyz}
M.~Dugger \textit{et al.} [CLAS],
Phys. Rev. C \textbf{79} (2009), 065206
[arXiv:0903.1110 [hep-ex]].


\bibitem{CLAS:2009ces}
I.~G.~Aznauryan \textit{et al.} [CLAS],
``Electroexcitation of nucleon resonances from CLAS data on single pion electroproduction,''
Phys. Rev. C \textbf{80} (2009), 055203
[arXiv:0909.2349 [nucl-ex]].


\bibitem{Mokeev2012}
V.~I.~Mokeev \textit{et al.} [CLAS],
``Experimental Study of the $P_{11}(1440)$ and $D_{13}(1520)$ resonances from CLAS data on $ep \rightarrow e'\pi^{+} \pi^{-} p'$,''
Phys. Rev. C \textbf{86}, 035203 (2012)
[arXiv:1205.3948 [nucl-ex]].

\bibitem{Mokeev2015}
V.~I.~Mokeev, V.~D.~Burkert, D.~S.~Carman, L.~Elouadrhiri, G.~V.~Fedotov, E.~N.~Golovatch, R.~W.~Gothe, K.~Hicks, B.~S.~Ishkhanov and E.~L.~Isupov, \textit{et al.}
``New Results from the Studies of the $N(1440)1/2^+$, $N(1520)3/2^-$, and $\Delta(1620)1/2^-$ Resonances in Exclusive $ep \to e'p' \pi^+ \pi^-$ Electroproduction with the CLAS Detector,'' Phys. Rev. C \textbf{93}, no.2, 025206 (2016)
[arXiv:1509.05460 [nucl-ex]].

\bibitem{Hata:2008xc}
H.~Hata, M.~Murata and S.~Yamato,
``Chiral currents and static properties of nucleons in holographic QCD,''
Phys. Rev. D \textbf{78}, 086006 (2008)
[arXiv:0803.0180 [hep-th]].

\bibitem{Burkert:2002zz}
V.~D.~Burkert, R.~De Vita, M.~Battaglieri, M.~Ripani and V.~Mokeev,
``Single quark transition model analysis of electromagnetic nucleon resonance transitions in the [70,1-] supermultiplet,''
Phys. Rev. C \textbf{67} (2003), 035204
[arXiv:hep-ph/0212108 [hep-ph]].

\bibitem{CLAS:2007bvs}
H.~Denizli \textit{et al.} [CLAS],
``Q*2 dependence of the S(11)(1535) photocoupling and evidence for a P-wave resonance in eta electroproduction,''
Phys. Rev. C \textbf{76} (2007), 015204
[arXiv:0704.2546 [nucl-ex]].

\bibitem{CLAS:2000mbw}
R.~Thompson \textit{et al.} [CLAS],
``The e p ---\ensuremath{>} e-prime p eta reaction at and above the S(11)(1535) baryon resonance,''
Phys. Rev. Lett. \textbf{86} (2001), 1702-1706
[arXiv:hep-ex/0011029 [hep-ex]].

\bibitem{Braaten:1986iw}
E.~Braaten, S.~M.~Tse and C.~Willcox,
``Electromagnetic Form-factors in the Skyrme Model,''
Phys. Rev. Lett. \textbf{56}, 2008 (1986)

\bibitem{A1:2008ocu}
S.~Stave \textit{et al.} [A1],
``Measurements of the gamma* p ---\ensuremath{>} Delta Reaction At Low Q**2: Probing the Mesonic Contribution,''
Phys. Rev. C \textbf{78} (2008), 025209
[arXiv:0803.2476 [hep-ex]].


\bibitem{OOPS:2004kai}
N.~F.~Sparveris \textit{et al.} [OOPS],
``Investigation of the conjectured nucleon deformation at low momentum transfer,''
Phys. Rev. Lett. \textbf{94} (2005), 022003
doi:10.1103/PhysRevLett.94.022003
[arXiv:nucl-ex/0408003 [nucl-ex]].

\bibitem{ParticleDataGroup:2020ssz}
P.~A.~Zyla \textit{et al.} [Particle Data Group],
``Review of Particle Physics,''
PTEP \textbf{2020} (2020) no.8, 083C01

\bibitem{HillerBlin:2022ltm}
A.~N.~Hiller Blin, V.~I.~Mokeev and W.~Melnitchouk,
``Resonant contributions to polarized proton structure functions,''
Phys. Rev. C \textbf{107} (2023) no.3, 035202
[arXiv:2212.11952 [hep-ph]].

\bibitem{inprogress}F. Castellani and A. Olzi, work in progress.

\bibitem{Kochelev:2011bh}
N.~Kochelev and Y.~Oh,
``Axial anomaly and the $\delta_{LT}$ puzzle,''
Phys. Rev. D \textbf{85}, 016012 (2012)
[arXiv:1103.4892 [hep-ph]].
 
\end{thebibliography}
\end{document}